\documentclass[aps,twocolumn,showpacs]{revtex4}
\usepackage{graphicx}
\usepackage{amsfonts}
\usepackage{amssymb}
\usepackage{amsbsy}
\usepackage{amsmath}
\usepackage{mathrsfs}
\usepackage{latexsym}
\usepackage{natbib}
\usepackage{bm}
\usepackage{subfigure} 
\usepackage{color}
\usepackage{wasysym}
\usepackage{mathbbol}
\usepackage{bigints}

\def\ca{c_{_{A}}}
\def\cb{c_{_{B}}}
\def\kp{k_{p}}

\begin{document}

\title{Role of thermal field in entanglement harvesting between two accelerated 
Unruh-DeWitt detectors}

\author{Dipankar Barman}
\email{dipankar1998@iitg.ac.in}

\author{Subhajit Barman}
\email{subhajit.b@iitg.ac.in}

\author{Bibhas Ranjan Majhi}
\email{bibhas.majhi@iitg.ac.in}

\affiliation{Department of Physics, Indian Institute of Technology Guwahati, 
Guwahati 781039, Assam, India}

\pacs{04.62.+v, 04.60.Pp}

\date{\today}

\begin{abstract}

We investigate the effects of field temperature $T^{(f)}$ on the entanglement 
harvesting between two uniformly accelerated detectors. For their parallel 
motion, the thermal nature of fields does not produce any entanglement, and 
therefore, the outcome is the same as the non-thermal situation. On the 
contrary, $T^{(f)}$ affects entanglement harvesting when the detectors are in 
anti-parallel motion, i.e., when detectors $A$ and $B$ are in the right and left 
Rindler wedges, respectively. While for $T^{(f)}=0$ entanglement harvesting is 
possible for all values of $A$'s acceleration $a_A$, in the presence of 
temperature, it is possible only within a narrow range of $a_A$. In $(1+1)$ 
dimensions, the range starts from specific values and extends to infinity, and 
as we increase $T^{(f)}$, the minimum required value of $a_A$ for entanglement 
harvesting increases. Moreover, above a critical value $a_A=a_c$ harvesting 
increases as we increase $T^{(f)}$, which is just opposite to the accelerations 
below it. There are several critical values in $(1+3)$ dimensions when they are 
in different accelerations. Contrary to the single range in $(1+1)$ dimensions, 
here harvesting is possible within several discrete ranges of $a_A$. 
Interestingly, for equal accelerations, one has a single critical point, with 
nature quite similar to $(1+1)$ dimensional results. We also discuss the 
dependence of mutual information among these detectors on $a_A$ and $T^{(f)}$.

\end{abstract}

\maketitle

\section{Introduction}\label{Introduction}

Quantum entanglement is a fascinating phenomenon distinguishing quantum and 
classical physics and has acquired immense practical importance through quantum 
communication and cryptography \cite{Tittel:1998ja, Salart-2008}. There has been 
a growing interest to realize entanglement and understand its nature for 
relativistic particles in flat and in curved spacetimes, see 
\cite{FuentesSchuller:2004xp, Reznik:2002fz, Lin:2010zzb, Ball:2005xa, 
Cliche:2009fma, MartinMartinez:2012sg, Salton:2014jaa, Martin-Martinez:2015qwa, 
Cai:2018xuo, Menezes:2017oeb, Menezes:2017rby, Zhou:2017axh, Floreanini:2004, 
Pan:2020tzf}. In this regard, entanglement extraction from the quantum field 
vacuum became very important for the fundamental understandings of the vacuum 
and the background spacetime. This phenomenon is better known as entanglement 
harvesting \cite{VALENTINI1991321, Reznik:2002fz, Reznik:2003mnx, 
Salton:2014jaa, Henderson:2017yuv, Henderson:2020ucx, Stritzelberger:2020hde}, 
which states that from quantum fields, one can harvest entanglement among atoms 
or other suitable systems interacting with the field. Entanglement harvesting 
acquires additional significance from the possibility of its experimental 
verification and utilization of the extracted entanglement in quantum 
information-related purposes \cite{Hotta:2008uk, Hotta:2009, Frey:2014}. In the 
pioneering works by Reznik \cite{Reznik:2002fz, Reznik:2003mnx}, he provided an 
understanding of entanglement harvesting in a system of two accelerated atoms by 
considering them as point-like two-level Unruh-DeWitt particle detectors 
\cite{2010grae.book} interacting with background massless scalar field.  
Unruh-DeWitt particle detectors are hypothetical detectors, conceptualized to 
understand the Unruh effect \cite{Unruh:1976db, Unruh:1983ms}. He showed that 
entanglement extraction was possible between two causally disconnected 
anti-parallelly accelerated detectors in the separate Rindler wedges, signifying 
the quantum vacuum's role for harvested entanglement.

There has been plenty of works related to entanglement harvesting in different 
spacetime backgrounds, and one can look into \cite{Martin-Martinez:2013eda, 
Lorek:2014dwa, VerSteeg:2007xs, Brown:2014pda, Pozas-Kerstjens:2015gta, 
Pozas-Kerstjens:2016rsh, Martin-Martinez:2015qwa, Kukita:2017etu, Sachs:2017exo, 
Trevison:2018ear, Li:2018xil} for a thorough anthology. In all these works, one 
usually investigates a system composed of two initially non-entangled detectors 
interacting with the background quantum field. The aim is to study the later 
time density matrix only for the detectors, where the field degrees of freedom 
are being traced out. For entanglement harvesting, i.e., for the two qubits to 
be entangled, it is necessary to have negative eigenvalues of the partial 
transposition of the detector density matrix. It should be noted that in initial 
works \cite{Reznik:2002fz, Salton:2014jaa} the authors found these eigenvalues 
to be dependent on quantities estimated from positive frequency Wightman 
functions connecting different spacetime events of the same or different 
detectors. However, recent rigorous investigations \cite{Koga:2018the, 
Ng:2018ilp, Koga:2019fqh} have suggested proper time ordering into the picture, 
which results in the inception of Feynman propagator rather than Wightman 
function in the estimation of the eigenvalues. These recent methods provide a 
meticulous and more general formulation for the understanding of entanglement 
harvesting. However, even with these changes, most previous perceptions 
regarding entanglement extraction corresponding to accelerated observers -- like 
one can harvest entanglement between two anti-parallelly accelerated detectors 
but not for parallelly accelerated observers -- remain the same. Although, the 
individual contributions of the retarded Green's function and the Wightman 
function from the Feynman propagator remain an interesting arena to venture 
further.

On the other hand, the effects of a thermal bath on entanglement harvesting 
remain equally interesting (see \cite{Brown:2013kia, Simidzija:2018ddw}). In 
nature, an environment with thermal background is much more practical. Including 
the thermal nature in the model and investigating the effects in the physical 
quantities will approach a more realistic situation and thereby help to know the 
exact features of our surroundings.  In this regard, one may mention that the 
thermal nature of fields has already been included in various investigations 
related to Unruh-De Witt detectors; like calculation of response functions in 
case of a single detector, \cite{Costa:1994yx, Kolekar:2013hra, 
Hodgkinson:2014iua, Chowdhury:2019set}, and two entangled detectors 
\cite{Barman:2021oum}. In \cite{Brown:2013kia, Simidzija:2018ddw, Lima:2020czr} 
it is predicted that the entanglement extraction gets depleted with increasing 
temperature of the thermal field. Then it will be pretty fascinating to study 
the situation of entanglement harvesting for accelerated observers interacting 
with thermal fields, which is not there in the literature up to our knowledge. 
In this regard, in literature \cite{Koga:2018the, Koga:2019fqh} the Feynman 
propagators and the positive frequency Wightman functions, necessary to 
understand entanglement extraction, are estimated in the Minkowski position 
space. Then for the calculations relating to accelerated observers, the relevant 
transformations to Rindler coordinates are made to those Green's functions, and 
this method does not encounter any particular issue. However, consider a similar 
description of the Feynman propagators and the Wightman functions for thermal 
fields. The resulting Green's functions do not remain time translational 
invariant with the detectors' proper times. In \cite{Barman:2021oum} the authors 
have discussed this issue and considered Rindler modes with the vacuum for the 
Unruh modes to describe the Wightman functions corresponding to accelerated 
observers in a thermal bath, which are time translational invariant. This 
method, in line with the chain of thoughts also presented in \cite{Ng:2018ilp}, 
circumvents the previously mentioned issue by expressing the Green's functions 
in terms of modes and their momentum space integrals rather than a position 
space representation.

In this work, we are going to investigate the condition for entanglement 
harvesting and study the concurrence \cite{Koga:2018the, Koga:2019fqh, 
Hu:2015lda}, a measure of the harvested entanglement, for two accelerated 
Unruh-DeWitt detectors interacting with a massless thermal scalar field in 
$(1+1)$ and $(1+3)$ dimensions. In particular, we consider the interaction 
between the two-level point-like detectors and the scalar field to be of 
monopole type. We observe that the specific form of this monopole moment 
operator is not needed to understand the role of the spacetime trajectories and 
the thermal bath in entanglement extraction. We use the prescription as provided 
in \cite{Barman:2021oum} for the construction of the Green's functions and 
follow the formulation of articles \cite{Koga:2018the, Koga:2019fqh} for 
entanglement harvesting. We arrive at the same assertions that entanglement 
extraction is possible only for the anti-parallelly accelerated detectors and 
not for the parallelly accelerated ones, and also encounter the phenomena of 
degrading entanglement extraction with increasing temperature of the thermal 
bath \cite{Brown:2013kia, Simidzija:2018ddw}. However, the situation is a bit 
more involved in our case as we observe this degradation happening in the low 
acceleration regimes. We observe that in $(1+1)$ dimensions, above a specific 
value of acceleration, thermal bath enhances the entanglement harvesting. While 
below this specific acceleration, the same is degraded with increasing 
background field temperature. Therefore, for anti-parallel detectors, a notion 
of phase transition-like phenomena is encountered around a critical acceleration 
value. However, we found that the range of acceleration in which entanglement 
harvesting is possible is consistently decreasing with the increasing 
temperature of the thermal bath. In $(1+3)$ dimensions, for equal accelerations 
of the detectors, the characteristics of concurrence are the same as the $(1+1)$ 
dimensional case.  However, for unequal detectors' accelerations, we encounter 
multiple transition points for $a_{A}$, the acceleration of detector $A$, when 
the acceleration $a_{B}$ of detector $B$ is fixed. We notice that between these 
transition points, the nature of the concurrence flips with the temperature of 
the thermal bath compared to the adjacent regions. In that case, contrary to the 
single range of $a_{A}$ in $(1+1)$ dimensions, we now have discrete ranges of 
acceleration $a_{A}$ for entanglement harvesting to be possible for a fixed 
temperature of the thermal bath. It is observed that a non-vanishing 
contribution is coming from the retarded part of the Feynman propagator when the 
detectors have unequal magnitudes of accelerations. We have also investigated 
the nature of mutual information $\mathcal{M}$ among the two detectors. Here 
$\mathcal{M}$ is non-vanishing for parallel motion, whereas it vanishes in the 
anti-parallel situation. For a non-vanishing case, $\mathcal{M}$ increases with 
the increase in temperature of the background field. On the other hand, it 
decreases with the growth of acceleration of the first detector.

In Sec. \ref{sec:radiative_process} we begin with a brief discussion of our 
model set-up of two two-level point-like atomic detectors interacting with the 
vacuum massless scalar field through monopole couplings. We consider the 
detectors initially in their separable ground state. This section also discusses 
the entanglement harvesting condition and entanglement measures obtained from 
the final form of the detector density matrix. In Sec. \ref{TR_spacetime} we 
elucidate on accelerated observers in a thermal bath and provide the expressions 
of the Green's functions for the situation of parallelly and anti-parallelly 
accelerating observers, considering the Rindler field decomposition with the 
Unruh operators and Unruh mode vacuum. Subsequently, in Sec. 
\ref{Entanglement-harvest} the condition for entanglement harvesting is analyzed 
first for two parallelly and then for two anti-parallelly accelerated observers 
in a thermal bath using the Green's functions of Sec. \ref{TR_spacetime}. In 
this section, we study the entanglement measure concurrence and, in Sec. 
\ref{Sec:MutualInf} investigate the mutual information between the two detectors 
to discuss the notable outcomes. We conclude this article in Sec. 
\ref{discussion} with a discussion of our results.

\section{Model set-up: a summary of the main 
results}\label{sec:radiative_process}

Having said our motivation in the introduction, let us now talk about the model 
which will be dealt with in this article. The model on which we will concentrate 
here was originally introduced in \cite{Koga:2018the, Ng:2018ilp, 
Koga:2019fqh}. Therefore, without going into the details of this and the 
derivation of the required formulas, the final expressions which are needed in 
this paper will be summarised here. Also a brief idea of the model will be given 
in order to be acquainted with the notations and symbols, we will use.

We consider two two-level point-like Unruh-DeWitt detectors, one carried by 
Alice and denoted by $A$. Another denoted by $B$, which is carried by Bob. The 
detector states are denoted by $|E_{n}^{j}\rangle$, with the symbols denoting 
the $n^{th}$ state of $j^{th}$ detector, i.e., $j=A,B$ and $n=0,1$. These states 
are non degenerate so that $E_{1}^{j}\neq E_{0}^{j}$, and it is assumed that 
$\Delta E^{j} = E_{1}^{j}- E_{0}^{j}>0$. We consider these detectors to be 
interacting through monopole interactions $m_{j}(\tau_{j})$ with a massless, 
minimally coupled scalar field $\Phi(X)$. The interaction action corresponding 
to this system is
\begin{eqnarray}\label{eq:TimeEvolution-int}
 S_{int} &=& \int_{-\infty}^{\infty} \bigg[ 
\ca\kappa_{A}(\tau_{A})m_{A}(\tau_{A}) \Phi\left(X_{ A}(\tau_{A})\right) 
d\tau_{A} \nonumber\\ 
&& +~ \cb \kappa_{B}(\tau_{B})m_{B}(\tau_{B}) \Phi\left(X_{ 
B}(\tau_{B})\right) d\tau_{B}\bigg]~,
\end{eqnarray}
where, $c_{j}$ denote the couplings between the individual detectors and the 
scalar field, $\kappa_{j}(\tau_{j})$ the switching functions and $\tau_{j}$ the 
individual detector proper times. The initial detector field state is considered 
to be the one at the asymptotic past, denoted by $|in\rangle = |0\rangle 
|E_{0}^{A}\rangle |E_{0}^{B}\rangle$. Whereas the final detector state at 
asymptotic future is $|out\rangle = T\left\{e^{i S_{int}}|in\rangle\right\}$. 
Treating the coupling constants $c_{_{j}}$ perturbatively and tracing out the 
field degrees of freedoms one can obtain the density matrix corresponding to the 
final state in the basis of $\big\{|E_{1}^{A}\rangle |E_{1}^{B}\rangle, 
|E_{1}^{A}\rangle |E_{0}^{B}\rangle, |E_{0}^{A}\rangle |E_{1}^{B}\rangle, 
|E_{0}^{A}\rangle |E_{0}^{B}\rangle\big\}$ as
\begin{widetext}
\begin{equation}\label{eq:detector-density-matrix}
 \rho_{AB} = 
 {\left[\begin{matrix}
 0 & 0 & 0 & \ca \cb\varepsilon\\~\\
0 & \ca^2P_{A} & \ca\cb P_{AB} & \ca^2 W_{A}^{(N)}+\ca\cb W_{A}^{(S)}\\~\\
0 & \ca\cb P_{AB}^{*} & \cb^2P_{B} & \cb^2 W_{B}^{(N)}+\ca\cb W_{B}^{(S)}\\~\\
\ca\cb\varepsilon^{*} & \ca^2 W_{A}^{(N){*}}+\ca\cb W_{A}^{(S){*}} & \cb^2 
W_{B}^{(N){*}}+\ca\cb W_{B}^{(S){*}} & 
1-(\ca^2 P_{A}+\cb^2 P_{B})
 \end{matrix}\right]}
 +\mathcal{O}(c^4)~,
\end{equation}
\end{widetext}
where, the expressions of $P_{j}$, $\varepsilon$, $P_{AB}$, 
$W_{j}^{(N)}$, and $W_{j}^{(S)}$ are given by
\begin{eqnarray}
P_{j} &=& |\langle E_{1}^{j}|m_{j}(0)|E_{0}^{j}\rangle|^2~
\mathcal{I}_{j}\nonumber
\end{eqnarray}
\begin{eqnarray}
\varepsilon &=& \langle E_{1}^{B}|m_{B}(0)|E_{0}^{B}\rangle\langle 
E_{1}^{A}|m_{A}(0)|E_{0}^{A}\rangle 
\mathcal{I}_{\varepsilon}\nonumber
\end{eqnarray}
\begin{eqnarray}
P_{AB} &=& \langle E_{1}^{A}|m_{A}(0)|E_{0}^{A}\rangle \langle 
E_{1}^{B}|m_{B}(0)|E_{0}^{B}\rangle^{\dagger} 
\mathcal{I}_{AB}\nonumber
\end{eqnarray}
\begin{eqnarray}
W_{j}^{(N)} &=& \langle E_{1}^{j}|m_{j}(0)|E_{0}^{j}\rangle\Big[\left(\langle 
E_{1}^{j}|m_{j}(0)|E_{1}^{j}\rangle -\right. \nonumber\\
~&& \left.\langle 
E_{0}^{j}|m_{j}(0)|E_{0}^{j}\rangle\right)
\mathcal{I}_{j,1}^{(N)}-i \langle 
E_{0}^{j}|m_{j}(0)|E_{0}^{j}\rangle
\mathcal{I}_{j,2}^{(N)}\Big]\nonumber
\end{eqnarray}
\begin{eqnarray}\label{eq:all-PJs}
W_{j}^{(S)} &=& -i\langle E_{1}^{j}|m_{j}(0)|E_{0}^{j}\rangle\langle 
E_{0}^{j'}|m_{j'}(0)|E_{0}^{j'}\rangle
\mathcal{I}_{j}^{(S)}~,
\end{eqnarray}
where $j'\neq j$ and the quantities $\mathcal{I}$,s are given by 
\begin{eqnarray}
 \mathcal{I}_{j} &=& \int_{-\infty}^{\infty}d\tau'_{j} 
\int_{-\infty}^{\infty}d\tau_{j}~e^{-i\Delta E^{j}(\tau'_{j}-\tau_{j})} 
G_{W}(X'_{j},X_{j}),\nonumber
\end{eqnarray}
\begin{eqnarray}
\mathcal{I}_{\varepsilon} &=& -i\int_{-\infty}^{\infty}d\tau'_{B} 
\int_{-\infty}^{\infty}d\tau_{A}~\scalebox{0.91}{$e^{i(\Delta 
E^{B}\tau'_{B}+\Delta E^{A}\tau_{A})} G_{F}(X'_{B},X_{A}),$}\nonumber
\end{eqnarray}
\begin{eqnarray}
\mathcal{I}_{AB} &=& \int_{-\infty}^{\infty}d\tau'_{B} 
\int_{-\infty}^{\infty}d\tau_{A}~\scalebox{0.91}{$e^{i(\Delta 
E^{A}\tau_{A}-\Delta E^{B}\tau'_{B})} G_{W}(X'_{B},X_{A}),$}\nonumber
\end{eqnarray}
\begin{eqnarray}
\mathcal{I}_{j,1}^{(N)} &=& \int_{-\infty}^{\infty}d\tau'_{j} 
\int_{-\infty}^{\infty}d\tau_{j}~\scalebox{0.91}{$e^{i\Delta 
E^{j}\tau_{j}}~ \theta(\tau'_{j}-\tau_{j})G_{W}(X'_{j},X_{j}),$}\nonumber
\end{eqnarray}
\begin{eqnarray}
\mathcal{I}_{j,2}^{(N)} &=& \int_{-\infty}^{\infty}d\tau'_{j} 
\int_{-\infty}^{\infty}d\tau_{j}~\scalebox{0.91}{$e^{i\Delta 
E^{j}\tau_{j}}~G_{R}(X_{j},X'_{j}),$}\nonumber
\end{eqnarray}
\begin{eqnarray}\label{eq:all-integrals}
\mathcal{I}_{j}^{(S)} &=& \int_{-\infty}^{\infty}d\tau'_{j'} 
\int_{-\infty}^{\infty}d\tau_{j}~\scalebox{0.91}{$e^{i\Delta 
E^{j}\tau_{j}}~G_{R}(X_{j},X'_{j'})~.$}
\end{eqnarray}
Here in these expressions the switching functions have not appeared as we have 
considered them $\kappa_{j}(\tau_{j})=1$; i.e. the detectors are interacting 
with fields all the time. On the other hand, the quantities 
$G_{W}(X_{j},X_{j'})$, $G_{F}(X_{j},X_{j'})$, and $G_{R}(X_{j},X_{j'})$ 
respectively denote the positive frequency Wightman function with 
$X_{j}>X_{j'}$, the Feynman propagator, and the retarded Green's function, and 
their expressions are
\begin{eqnarray}\label{eq:Greens-fn-gen}
 G_{W}\left(X_{j},X_{j'}\right) &\equiv& \langle 
0_{M}|\Phi\left(X_{j}\right)\Phi\left(X_{j'}\right)|0_{M}
\rangle~,\nonumber\\
 G_{F}\left(X_{j},X_{j'}\right) &\equiv& -i\langle 
0_{M}|T\left\{\Phi\left(X_{j}\right)\Phi\left(X_{j'}\right)\right\}|0_{M}
\rangle~,\nonumber\\
 G_{R}\left(X_{j},X_{j'}\right) &\equiv& i\theta(t-t')\langle 
0_{M}|\left[\Phi\left(X_{j'}\right),\Phi\left(X_{j}\right)\right]|0_{M}
\rangle.\nonumber\\
\end{eqnarray}
The details of the derivation can be followed from \cite{Koga:2018the}. The 
condition for entanglement, based on a general analysis for a bipartite system 
\cite{Peres:1996dw, Horodecki:1996nc}, is obtained from the negative eigenvalue 
of the partial transposition of the reduced density matrix from Eq. 
(\ref{eq:detector-density-matrix}), and this condition results in
\begin{equation}
 P_{A}P_{B}<|\varepsilon|^2~, 
\end{equation}
which can also be cast, in terms of the integrals, into the form 
\cite{Koga:2018the, Koga:2019fqh}
\begin{equation}\label{eq:cond-entanglement}
 \mathcal{I}_{A}\mathcal{I}_{B}<|\mathcal{I}_{\varepsilon}|^2~.
\end{equation}
Now one can use the relation between Feynman propagator and the Wightman 
function $iG_{F}\left(X_{j},X_{j'}\right) = G_{W}\left(X_{j},X_{j'}\right) + i 
G_{R}\left(X_{j'},X_{j}\right) = G_{W}\left(X_{j},X_{j'}\right) + \theta(T'-T) 
\left\{G_{W}\left(X_{j'},X_{j}\right)-G_{W}\left(X_{j},X_{j'}\right)\right\}$ 
to simplify the calculation of the integral $\mathcal{I}_{\varepsilon}$. 
In particular, one can now express that integral as 
\begin{eqnarray}\label{eq:Ie-integral}
 && \mathcal{I}_{\varepsilon} = -\int_{-\infty}^{\infty}d\tau_{B} 
\int_{-\infty}^{\infty}d\tau_{A}~\scalebox{0.91}{$e^{i(\Delta 
E^{B}\tau_{B}+\Delta E^{A}\tau_{A})} $}\times\nonumber\\
~&& \scalebox{0.87}{$\left[G_{W}(X_{B},X_{A})+ \theta(T_{A}-T_{B}) 
\left\{G_{W}\left(X_{A},X_{B}\right)-G_{W}\left(X_{B},X_{A}\right)\right\}
\right].$}\nonumber\\
\end{eqnarray}
We will use the above form for our purpose. It is observed that one only needs 
the expressions of the integrals $\mathcal{I}_{A}$, $\mathcal{I}_{B}$  and 
$\mathcal{I}_{\varepsilon}$ for verification of the condition 
(\ref{eq:cond-entanglement}) for entanglement harvesting. From Eq. 
(\ref{eq:all-integrals}) and (\ref{eq:Ie-integral}) we observe that all of 
these 
integrals can be evaluated in terms of the positive frequency Wightman 
functions.

When the condition for entanglement harvesting (\ref{eq:cond-entanglement}) is 
satisfied, it is convenient to study different entanglement measures. In this 
regard, one relevant entanglement measure is the negativity 
\cite{Zyczkowski:1998yd, Vidal:2002zz, Eisert:1998pz, Devetak_2005}, which 
signifies the upper bound of the distillable entanglement and is obtained from 
the sum of all negative eigenvalues of the partial transpose of $\rho_{AB}$. In 
the two qubits case another important and more convenient entanglement measure 
is the concurrence $\mathcal{C}(\rho_{AB})$ \cite{Koga:2018the, Koga:2019fqh, 
Hu:2015lda}, from which entanglement of formation $E_{F}(\rho_{AB})$ is 
estimated, see \cite{Bennett:1996gf, Hill:1997pfa, Wootters:1997id, 
Koga:2018the, Koga:2019fqh}. For two-qubits, the concurrence is given by, see 
\cite{Koga:2018the},
\begin{eqnarray}\label{eq:concurrence-gen-exp}
 \mathcal{C}(\rho_{AB}) &=& max\bigg[0,~ 2c^2 
\left(|\varepsilon|-\sqrt{P_{A}P_{B}}\right)+\mathcal{O}(c^4)\bigg]\nonumber\\
~&=& max\bigg[0,~2c^2|\langle E_{1}^{B}|m_{B}(0)| E_{0}^{B}\rangle| |\langle 
E_{1}^{A}|m_{A}(0)| E_{0}^{A}\rangle|\nonumber\\
~&& ~~~~~~\times 
\left(|\mathcal{I}_{\varepsilon}|-\sqrt{\mathcal{I}_{A}\mathcal{I}_{B}}
\right)+\mathcal{O} (c^4)\bigg]~,
\end{eqnarray}
where, an equal magnitude of the coupling constant $c_{A}=c_{B}=c$ between 
different detectors and the scalar field is assumed. It should be noted that 
the 
quantities $|\langle E_{1}^{j}|m_{j}(0)| E_{0}^{j}\rangle|$ are specified by 
the 
detectors' internal structure and do not take contributions from the considered 
spacetime and background scalar fields.  Since we are interested to investigate 
the entanglement harvesting due to the motions of these detectors, then for a 
specific detector configuration it is only relevant to study the nature of 
\begin{equation}\label{eq:concurrence-I}
\mathcal{C}_{\mathcal{I}} = \left(|\mathcal{I} 
_{\varepsilon}| -\sqrt{\mathcal{I}_{A}\mathcal{I}_{B}} \right)
\end{equation} 
as far as concurrence is concerned. It should also be noted that in the 
symmetric case $\mathcal{I}_{A}=\mathcal{I}_{B}$, which for example, can happen 
in the case of the equal magnitude of the acceleration of the two detectors, 
this relevant quantity signifying the concurrence is given by 
$\mathcal{C}_{\mathcal{I}} =  \left(|\mathcal{I}_{\varepsilon}| 
-\mathcal{I}_{j}\right)$, see \cite{Koga:2018the, Koga:2019fqh}. In our later 
analysis, we shall be studying this $\mathcal{C}_{\mathcal{I}}$ to talk about 
the entanglement measure in our considered system. Particularly, by this, we 
will be investing the nature of entanglement harvesting for different parameters 
of our system.

On the other hand, the total correlations, i.e., the entirety of classical and 
quantum correlations, between the two detectors $A$ and $B$ with the observers 
Alice and Bob is quantified by mutual information $\mathcal{M}$, defined as 
\begin{equation}\label{eq:MI-general}
 \mathcal{M}(\rho_{AB}) \equiv S(\rho_{A}) + S(\rho_{B}) -S(\rho_{AB})~,
\end{equation}
where, $\rho_{A}\equiv Tr_{B}(\rho_{AB})$ and $\rho_{B}\equiv Tr_{A}(\rho_{AB})$ 
are the reduced density matrices corresponding to the detectors $A$ and $B$, and 
$S(\rho)\equiv -Tr(\rho\ln{\rho})$ is the von Neumann entropy corresponding to 
the state with $\rho$ to be the density matrix. Using the expression of the 
density matrix from Eq. (\ref{eq:detector-density-matrix}), and considering the 
equal couplings between the field and the two detectors, one can express the 
mutual information of (\ref{eq:MI-general}) as \cite{Simidzija:2018ddw}
\begin{eqnarray}\label{eq:MI-explicit}
 \mathcal{M}(\rho_{AB}) &=& c^2\big[P_{+}\ln{P_{+}} + P_{-}\ln{P_{-}} - 
P_{A}\ln{P_{A}}\nonumber\\
~&&~~~~~~~~ -~ P_{B}\ln{P_{B}}\big] + \mathcal{O}(c^4)~,
\end{eqnarray}
where, the quantities $P_{\pm}$ are given by
\begin{equation}\label{eq:P-pm}
 P_{\pm} = \frac{1}{2} \Big[ P_{A}+P_{B}\pm \sqrt{(P_{A}-P_{B})^2+4|P_{AB}|^2} 
\Big]~.
\end{equation}
We mention that one may encounter situations when both the concurrence and 
mutual information are not simultaneously non-zero for a specific system. 
Between the concurrence and mutual information, if only the latter is non-zero, 
the correlation is considered classical. Therefore, it is necessary to 
investigate both of these measures to understand the correlation between the two 
detectors.

\section{Accelerated observers in a thermal bath}\label{TR_spacetime}

This section discusses the relevant coordinate systems for our accelerated 
observers. We realized that the whole analysis is more convenient under the 
decomposition of field modes in the Rindler frame and writing the Rindler 
annihilation and creation operators in terms of those of Unruh modes'. This 
will 
be introduced in a separate subsection. Finally, all the required positive 
frequency Wightman functions, both in $(1+1)$ and $(1+3)$ dimensions, for these 
fields will be evaluated with respect to the  Minkowski vacuum, which is also 
the vacuum for Unruh modes.

\subsection{Coordinate systems}\label{Rindler-coordinates}

The motion of a uniformly accelerated object is described by the Rindler 
coordinates which correspond to specific regions in the Minkowski spacetime, 
known as the Rindler wedges \cite{Crispino:2007eb}. One can move to these 
Rindler coordinates from the flat Minkowski coordinates $(T,X,Y,Z)$ in $(3+1)$ 
dimensions, with the line element
\begin{equation}\label{eq:Minkowski-metric}
 ds^2 = -dT^2+dX^2+dY^2+dZ^2~,
\end{equation}
by a coordinate transformation relating the time $T$ and the spatial direction 
in which the object is accelerated. Without loss of generality we consider that 
particular axis of acceleration to be along the Minkowski $X$ direction. Then 
the other two coordinates $(Y,Z)$ remain unchanged by the Rindler 
transformation. The transformations to the coordinates $(\eta,\xi)$ in the 
right 
Rindler wedge (RRW), i.e., the region $|T|<X$ in the Minkowski spacetime; and 
to 
$(\eta',\xi')$ in the left Rindler wedge (LRW), confined in a region $|T|<-X$ 
of 
the Minkowski spacetime, are
\begin{eqnarray}\label{eq:Rindler1-trans1}
 T &=& \frac{e^{a\xi}}{a} \sinh{a\eta},~ X = 
\frac{e^{a\xi}}{a} \cosh{a\eta}~~~~~~~~~\textup{in RRW}~;\nonumber\\
 T &=& -\frac{e^{a\xi'}}{a} \sinh{a\eta'},~
 X = -\frac{e^{a\xi'}}{a} \cosh{a\eta'}~\textup{in LRW}.
\end{eqnarray}
Both of these transformations lead to the same line-element corresponding to an 
accelerated observer in terms of the Rindler coordinates, expressed as
\begin{equation}\label{eq:Rindler1-metric}
 ds^2 = e^{2a\xi}\left[-d\eta^2+d\xi^2\right]+dY^2+dZ^2~.
\end{equation}
One can perceive that these transformations in $(1+1)$ dimensions are trivially 
same as in that case the coordinates $Y$ and $Z$ cease to exist. In RRW and LRW 
one can estimate the proper times and proper accelerations to be 
\begin{eqnarray}\label{eq:RindProper-time-acceleration}
&& \tau=e^{a\xi}\eta, ~ b=ae^{-a\xi}~~~~~~~~~~~~\textup{in RRW}~;\nonumber\\
&& \tau'=-e^{a\xi'}\eta', ~ b'=ae^{-a\xi'}~~~~~~\textup{in LRW}~.
\end{eqnarray}
Then the coordinate transformations (\ref{eq:Rindler1-trans1}) in terms of 
proper time and acceleration become
\begin{eqnarray}\label{eq:Rindler1-trans2}
 T &=& \frac{1}{b} \sinh{b\tau},~
 X = \frac{1}{b} \cosh{b\tau}~~~~~~~~~\textup{in RRW};\nonumber\\
  T &=& \frac{1}{b'} \sinh{b'\tau'},~
 X = -\frac{1}{b'} \cosh{b'\tau'}~~\textup{in LRW}.
\end{eqnarray}
One can notice that $\eta$, $-\eta'$ denote the proper times in RRW and LRW 
respectively while $a$ is the proper acceleration of the observer when 
$\xi=0=\xi'$.

\subsection{Scalar field decomposition corresponding to an accelerated observer}

To address the situation of an accelerated observer in a thermal bath one can 
consider expressing the thermal two-point function in terms of the Minkowski 
modes and then make the Rindler coordinate transformation from Eq. 
(\ref{eq:Rindler1-trans2}). However, the Green's function obtained in this way, 
for thermal field, is not time translational invariant in terms of proper time 
and a prescription to obtain unit time detector response using them is not 
possible \cite{Barman:2021oum}. On the other hand, one can also express the 
scalar field $\Phi(x)$ in terms of the Rindler modes and operators for which 
the corresponding vacuum is the Rindler vacuum. Then using the procedure as 
presented by Unruh in $1976$ (see \cite{Unruh:1976db}), by transforming the 
Rindler operators to the Unruh operators which correspond to the vacuum of the 
Unruh modes (which is here Minkowski vacuum), one can construct Wightman 
function corresponding to accelerating observers in thermal Minkowski 
background. This way of construction provides the proper time translation 
invariance in a natural way and analysis becomes analytically more tractable 
(e.g. see \cite{Barman:2021oum}).

The procedure of decomposing the scalar field in terms of the Unruh operators 
is 
elaborately discussed in \cite{book:Birrell, book:carroll}. Here we give a 
brief 
recollection of the construction and refer to the article \cite{Barman:2021oum} 
for further understandings. We first consider the case in $(1+1)$ dimensions 
and 
the $(1+3)$ dimensional result will follow accordingly. The equation of motion 
for a minimally coupled, massless free scalar field $\Phi$ is expressed as 
$\Box\Phi=0$. 

\subsubsection{$(1+1)$ dimensions}
In terms of the Rindler coordinates in $(1+1)$ dimensions this equation has 
solutions, suggesting set of modes in the right and left Rindler wedges as 
\cite{book:Birrell, book:carroll}
\begin{eqnarray}\label{eq:Rindler-modes}
 ^{R}u_{k} &=& \frac{1}{\sqrt{4\pi\omega}} 
e^{ik\xi-i\omega\eta}~~~\textup{in~RRW}\nonumber\\ 
~&=& 0~~~~~~~~~~~~~~~~~~~~\textup{in~LRW}\nonumber\\
 ^{L}u_{k} &=& \frac{1}{\sqrt{4\pi\omega}} 
e^{ik\xi+i\omega\eta}~~~\textup{in~LRW}\nonumber\\
~&=& 0~~~~~~~~~~~~~~~~~~~~\textup{in~RRW}.
\end{eqnarray}
The scalar field is expressed in terms of the Rindler modes and operators, see 
\cite{book:Birrell}, as $\Phi(X) = \sum_{k=-\infty}^{\infty} \left[b^{R}_{k}~ 
^{R}u_{k} + b^{R^{\dagger}}_{k}~ ^{R}u_{k}^{*}+b^{L}_{k}~ ^{L}u_{k} + 
b^{L^{\dagger}}_{k}~ ^{L}u_{k}^{*}\right]$, where superscript $L$ and $R$ 
correspond to the left and the right Rindler wedges respectively, and the 
annihilation operators annihilate the Rindler vacuum $|0_{\mathcal{R}}\rangle$, 
i.e. $b_{k}^{R}|0_{\mathcal{R}}\rangle =0 =b_{k}^{L}|0_{\mathcal{R}}\rangle$. 
In 
the right or left Rindler wedges where the field modes $ ^{L}u_{k}=0$ or $ 
^{R}u_{k}=0$, the scalar field takes the form
\begin{eqnarray}\label{eq:Field-Rindler-1p1}
 \Phi^{R}(X) &=& \sum_{k=-\infty}^{\infty} 
\left[b^{R}_{k}~ ^{R}u_{k} + 
b^{R^{\dagger}}_{k}~ ^{R}u_{k}^{*}\right]~,\nonumber\\
\textup{or}~~~\Phi^{L}(X) &=& \sum_{k=-\infty}^{\infty} 
\left[b^{L}_{k}~ ^{L}u_{k} + 
b^{L^{\dagger}}_{k}~ ^{L}u_{k}^{*}\right] ~.
\end{eqnarray}
One can use this scalar field decomposition in terms of the Rindler modes and 
operators to obtain a two-point function corresponding to an accelerated 
observer in Minkowski vacuum. Here, it should be noted that the operators 
$b^{R}_{k}$ and $b^{L}_{k}$ in Eq. (\ref{eq:Field-Rindler-1p1})  do not 
annihilate the Minkowski vacuum, and the operations of the Rindler ladder 
operators on the Minkowski vacuum is obtained from the cumbersome calculations 
of Bogoliubov transformation. However, there is a simpler way out of this 
situation as provided by Unruh \cite{Unruh:1976db}, where he prescribed field 
modes out of these Rindler modes which are analytic in the whole region of the 
Minkowski spacetime. These Unruh modes have the positive frequency analyticity 
property with respect to the Minkowski time, same as the Minkowski modes. This 
enables one to decompose the scalar field in terms of these Unruh modes and 
operators, which annihilate the Minkowski vacuum. The Unruh modes are obtained 
from the combination of the Rindler modes $^{R}u_{k} + e^{-\pi\omega/a}~ 
^{L}u^{*}_{-k}$ and $^{R}u^{*}_{-k} + e^{\pi\omega/a}~ ^{L}u_{k}$, see 
\cite{book:Birrell}. In terms of the Unruh modes and operators the scalar field 
is expressed as \cite{book:Birrell}
\begin{eqnarray}\label{eq:Field-Unruh}
 \Phi(X) &=& \sum_{k=-\infty}^{\infty} 
\frac{1}{\sqrt{2\sinh{\frac{\pi\omega}{a}}}} 
\left[d^{1}_{k}\left(e^{\frac{\pi\omega}{2a}}~ ^{R}u_{k} + 
e^{-\frac{\pi\omega}{2a}}~ ^{L}u^{*}_{-k}\right)\right.\nonumber\\
~&& ~~~+~ \left. d^{2}_{k}\left(e^{-\frac{\pi\omega}{2a}}~ ^{R}u^{*}_{-k} + 
e^{\frac{\pi\omega}{2a}}~ ^{L}u_{k}\right) \right] + h.c.~,
\end{eqnarray}
where $h.c.$ stands for Hermitian conjugate. The Unruh annihilation operators 
annihilate the Minkowski vacuum $ d^{1}_{k}|0_{M}\rangle = 
d^{2}_{k}|0_{M}\rangle=0~$. To obtain the positive frequency Green's function 
using the field decompositions of Eq. (\ref{eq:Field-Rindler-1p1}), one needs a 
transformation between the Rindler operators and the Unruh operators, see 
\cite{book:Birrell}, which is
\begin{eqnarray}\label{eq:bR-1p1-RRW}
b_{k}^{L} &=& \frac{1}{\sqrt{2\sinh{\frac{\pi\omega}{a}}}} 
\left[e^{\frac{\pi\omega}{2a}} d_{k}^{2}+e^{-\frac{\pi\omega}{2a}} 
d_{-k}^{1^{\dagger}}\right]\nonumber\\
 b_{k}^{R} &=& \frac{1}{\sqrt{2\sinh{\frac{\pi\omega}{a}}}} 
\left[e^{\frac{\pi\omega}{2a}} d_{k}^{1}+e^{-\frac{\pi\omega}{2a}} 
d_{-k}^{2^{\dagger}}\right]~,
\end{eqnarray}
and, it is similar to the Bogoliubov transformation. Then putting this 
transformation in Eq. (\ref{eq:Field-Rindler-1p1}) one can get the 
expression of the field in the RRW and LRW in terms of the Unruh operators as
\begin{eqnarray}\label{eq:RRW-LRW-field-Unruh}
 \Phi^{R}(X) &=& \sum_{k=-\infty}^{\infty} 
\frac{1}{\sqrt{2\sinh{\frac{\pi\omega}{a}}}} 
\left[d^{1}_{k}~e^{\frac{\pi\omega}{2a}}~ ^{R}u_{k} \right.\nonumber\\
~&& ~~~+~ \left. d^{2}_{k}~e^{-\frac{\pi\omega}{2a}}~ ^{R}u^{*}_{-k} \right] + 
h.c.~,\nonumber\\
\Phi^{L}(X) &=& \sum_{k=-\infty}^{\infty} 
\frac{1}{\sqrt{2\sinh{\frac{\pi\omega}{a}}}} 
\left[d^{1}_{k} 
e^{-\frac{\pi\omega}{2a}}~ ^{L}u^{*}_{-k}\right.\nonumber\\
~&& ~~~+~ \left. d^{2}_{k} 
e^{\frac{\pi\omega}{2a}}~ ^{L}u_{k} \right] + 
h.c.~.
\end{eqnarray}
Now these expression of the scalar fields in RRW and LRW can be used to obtain 
the expressions of the positive frequency Green's function corresponding to 
accelerated observers in thermal bath.

\subsubsection{$(1+3)$ dimensions}
Like the above analysis, in $(1+3)$ dimensions also, one can proceed in a 
similar 
manner to get the Scalar field in terms of the Unruh operators. In particular, 
from the equation of motion $\Box\Phi=0$ one can get the Rindler modes in the 
right and the left Rindler wedges as
\begin{eqnarray}\label{eq:Rindler-modes-3D}
 ^{R}u_{\omega,\kp} &=& 
\frac{1}{2\pi^2}\sqrt{\frac{\sinh{\left(\frac{\pi\omega}{a}\right)}}{a}}~ 
\mathcal{K}\left[\frac{i\omega}{a},\frac{|\kp| e^{a\xi}}{a}\right]\nonumber\\
~&& ~~~~~~\times~~ 
e^{-i\omega\eta+i\vec{\kp}.\vec{x}}~~~~~~~~\textup{in~RRW}\nonumber\\ 
~&=& 0~~~~~~~~~~~~~~~~~~~~~~~~~~~~~~~~~\textup{in~LRW}\nonumber\\
 ^{L}u_{\omega,\kp} &=& 
\frac{1}{2\pi^2}\sqrt{\frac{\sinh{\left(\frac{\pi\omega}{a}\right)}}{a}}~ 
\mathcal{K}\left[\frac{i\omega}{a},\frac{|\kp| e^{a\xi}}{a}\right]\nonumber\\
 ~&&~~~~~~\times~~
e^{i\omega\eta+i\vec{\kp}.\vec{x}}~~~~~~~~~~\textup{in~LRW}\nonumber\\
~&=& 0~~~~~~~~~~~~~~~~~~~~~~~~~~~~~~~~~\textup{in~RRW}~,
\end{eqnarray}
where, $\mathcal{K}\left[n,z\right]$ denotes the modified Bessel function of 
the 
second kind of order $n$, $\vec{x}$ is perpendicular to the direction of 
acceleration, i.e., in the $Y-Z$ plane, see \cite{Compere:2019rof, 
Crispino:2007eb, Higuchi:2017gcd}, and $\vec{\kp}$ denotes the transverse wave 
vector in the $Y-Z$ plane. Like the $(1+1)$ dimensional case here also one can 
construct the Unruh modes \cite{Crispino:2007eb} out of the Rindler modes, 
which 
are analytic in the whole Minkowski spacetime and gives positive frequency mode 
solutions with respect to the Minkowski time. Then in $(1+3)$ dimensions the 
scalar field in the RRW and LRW using the Unruh operators, see 
\cite{Barman:2021oum, Crispino:2007eb} for a detailed description, can be 
expressed in forms 
\begin{eqnarray}\label{eq:RRW-field-Unruh-1p3}
 \Phi^{R}(X) = \sum_{\omega=0}^{\infty}\sum_{\kp=-\infty}^{\infty} 
\frac{1}{\sqrt{2\sinh{\frac{\pi\omega}{a}}}} \times 
~~~~~~~~~~~~~~~~~~~~~&&\nonumber\\
\left[d^{1}_{\omega,\kp}e^{\frac{\pi\omega}{2a}}~ ^{R}u_{\omega,\kp} 
 + d^{2}_{\omega,\kp}e^{-\frac{\pi\omega}{2a}}~ 
^{R}u^{*}_{\omega,-\kp} \right] + h.c.~,&&\nonumber\\
\Phi^{L}(X) = \sum_{\omega=0}^{\infty}\sum_{\kp=-\infty}^{\infty} 
\frac{1}{\sqrt{2\sinh{\frac{\pi\omega}{a}}}} \times 
~~~~~~~~~~~~~~~~~~~~~&&\nonumber\\
\left[d^{1}_{\omega,\kp}e^{-\frac{\pi\omega}{2a}}~ ^{L}u^{*}_{\omega,-\kp} 
 + d^{2}_{\omega,\kp}e^{\frac{\pi\omega}{2a}}~ 
^{L}u_{\omega,\kp} \right] + h.c.~.&&\nonumber\\
\end{eqnarray}
This is exactly same as the $(1+1)$ dimensional expression with the Rindler 
field modes $^{R}u_{\omega,\kp}$ and $^{L}u_{\omega,\kp}$ are now given by 
different expressions, and the sum is now on $\omega$ and two components of 
$\kp$ rather than one wave vector $k$ of the $(1+1)$ dimensional case.

\subsection{Two-point correlators for thermal field}\label{TwoPointFn-thermal}

Considering a scalar field $\Phi(X)=\Phi(T,\mathbf{X})$ in equilibrium with a 
thermal bath of temperature $T^{(f)}=1/(k_{B}\beta)$, where $k_{B}$ is the 
\emph{Boltzmann constant}, the thermal Green's (Wightman) function can be 
obtained by taking Gibbs ensemble average of the operator 
$\Phi(X_{2})\Phi(X_{1})$ as 
\begin{eqnarray}\label{eq:Greens-fn-thermal1}
 G^{\beta}(X_{2};X_{1}) &=& \langle \Phi(X_{2})\Phi(X_{1}) 
\rangle_{\beta}\nonumber\\
 ~&=& \frac{1}{Z}~\textrm{Tr}\left[e^{-\beta H} \Phi(X_{2})\Phi(X_{1})\right] ~,
\end{eqnarray}
where, $X_{1}$ and $X_{2}$ are two events in the spacetime,  $Z = 
\textrm{Tr}[\exp(-\beta H)]$ denotes the partition function, and $H$ denotes 
the 
Hamiltonian of free massless scalar field.

\subsubsection{$(1+1)$-dimensions}

In $(1+1)$ dimensions to obtain the thermal Green's function corresponding to 
accelerated observers, with respect to Rindler modes, we consider massless 
scalar field where $\omega=\omega_{k}=|k|$. The Hamiltonian related to the 
$k^{th}$ excitation corresponding to the Unruh operators, which respect the 
Unruh vacuum, is $H_{k} = (d^{1^{\dagger}}_{k} d^{1}_{k}+ 
d^{2^{\dagger}}_{k}d^{2}_{k})\omega_{k}$. Then the thermal Green's function, 
defined by Eq. (\ref{eq:Greens-fn-thermal1}), corresponding to an accelerated 
observer, see \cite{Barman:2021oum}, can be expressed as
\begin{eqnarray}\label{eq:Greens-fn-TU-RRW}
&& G^{\beta}_{W_{R}}\left(\Delta\xi_{jl},\Delta\eta_{jl}\right)\nonumber\\  
&&~~~~~~= \int_{-\infty}^{\infty} 
\frac{dk}{8\pi\omega_{k}\sqrt{\sinh{\frac{\pi\omega_{k}}{a_{j}}}\sinh{
\frac{\pi\omega_{k}}{a_{l}}}}}\times\nonumber\\ 
&&~~~~~~\left[\frac{1}{1-e^{-\beta\omega_{k}
}}\left\{e^{ik\Delta\xi_{jl}-i\omega_{k}\Delta\eta_{jl}} 
~e^{\frac{\pi\omega_{k}}{2}\left(\frac{1}{a_{j}}+\frac{1}{a_{l}}\right)}
\right.\right. \nonumber\\
~&&~~~~~~ \left. ~~+~ 
e^{ik\Delta\xi_{jl}+i\omega_{k}\Delta\eta_{jl}}~
e^{-\frac{\pi\omega_{k}}{2}\left(\frac{1}{a_{j}}+\frac{1}{a_{l}}\right)}
\right\}\nonumber\\
&&~~~~~~ +  \frac{1}{e^{ \beta\omega_{k}}-1}
\left\{e^{-ik\Delta\xi_{jl}+i\omega_{k}\Delta\eta_{jl}} 
~e^{\frac{\pi\omega_{k}}{2}\left(\frac{1}{a_{j}}+\frac{1}{a_{l}}\right)}
\right.\nonumber\\ 
~&&~~~~~~ \left.\left.~~ +~ e^{-ik\Delta\xi_{jl}-i\omega_{k}\Delta\eta_{jl}}~ 
e^{-\frac{\pi\omega_{k}}{2}\left(\frac{1}{a_{j}}+\frac{1}{a_{l}}\right)} 
\right\}\right ] , 
\end{eqnarray}
where, $j$, $l$ denote different detectors, and 
$\Delta\xi_{jl}=\xi_{j,2}-\xi_{l,1}$, $\Delta\eta_{jl}= \eta_{j,2} 
-\eta_{l,1}$. 
For observers in the left Rindler wedge immersed in a thermal bath the Wightman 
function $G^{\beta}_{W_{L}}\left(\Delta\xi_{jl}, \Delta\eta_{jl}\right)$ is 
obtained from the expression of Eq. (\ref{eq:Greens-fn-TU-RRW}) with 
$\Delta\eta_{jl}\to-\Delta\eta_{jl}$.
%
%

Similarly for observers with one in the right Rindler wedge and one in the left 
Rindler wedge immersed in a thermal bath the Wightman function can be 
expressed, using the appropriate field modes from Eq. 
(\ref{eq:RRW-LRW-field-Unruh}), as
\begin{eqnarray}\label{eq:Greens-fn-TU-LR}
&& G^{\beta}_{W_{LR}}\left(\Delta\xi_{jl},\Delta\eta_{jl}\right)\nonumber\\ 
&&~~~~~~= \int_{-\infty}^{\infty} 
\frac{dk}{8\pi\omega_{k}\sqrt{\sinh{\frac{\pi\omega_{k}}{a_{j}}}\sinh{
\frac{\pi\omega_{k}}{a_{l}}}}}\times\nonumber\\ 
&&~~~~~~\left[\frac{1}{1-e^{-\beta\omega_{k}
}}\left\{e^{ik\Delta\xi_{jl}-i\omega_{k}\Delta\eta_{jl}} 
~e^{-\frac{\pi\omega_{k}}{2}\left(\frac{1}{a_{j}}-\frac{1}{a_{l}}\right)}
\right.\right. \nonumber\\
~&&~~~~~~ \left. ~~+~ 
e^{ik\Delta\xi_{jl}+i\omega_{k}\Delta\eta_{jl}}~
e^{\frac{\pi\omega_{k}}{2}\left(\frac{1}{a_{j}}-\frac{1}{a_{l}}\right)}
\right\}\nonumber\\
&&~~~~~~ +  \frac{1}{e^{ \beta\omega_{k}}-1}
\left\{e^{-ik\Delta\xi_{jl}+i\omega_{k}\Delta\eta_{jl}} 
~e^{-\frac{\pi\omega_{k}}{2}\left(\frac{1}{a_{j}}-\frac{1}{a_{l}}\right)}
\right.\nonumber\\ 
~&&~~~~~~ \left.\left.~~ +~ e^{-ik\Delta\xi_{jl}-i\omega_{k}\Delta\eta_{jl}}~ 
e^{\frac{\pi\omega_{k}}{2}\left(\frac{1}{a_{j}}-\frac{1}{a_{l}}\right)} 
\right\}\right ] , 
\end{eqnarray}
where, we have considered the $j^{th}$ detector to be in the left Rindler wedge 
and the detector denoted by $l$ is in the right Rindler wedge. We also mention 
that the Wightman function $G_{W_{RL}}^{\beta} \left(\Delta\xi_{jl}, 
\Delta\eta_{jl}\right)$, where the detectors denoted by $j$ and $l$ are in 
right 
and left Rindler wedges is obtained from the complex conjugate of the 
expression 
in the right hand side of Eq. (\ref{eq:Greens-fn-TU-LR}).
It should be noted that from Eq. (\ref{eq:Greens-fn-TU-RRW}) the thermal 
Green's 
function corresponding to a single accelerated detector can also be obtained by 
making $a_{j}=a_{l}$.

\subsubsection{$(1+3)$ dimensions}

One can obtain the thermal Green's function corresponding to accelerated 
observers, with respect to Rindler modes in $(1+3)$ dimensions in a similar 
manner. The field decomposition is taken from Eq. 
(\ref{eq:RRW-field-Unruh-1p3}) 
and the Hamiltonian corresponding to the Unruh operators is $H_{\omega,\kp} = 
(d^{1^{\dagger}}_{\omega,\kp} d^{1}_{\omega,\kp} + d^{2^{\dagger}}_{\omega,\kp} 
d^{2}_{\omega,\kp})\omega$. Then in RRW the Green's function corresponding to 
an accelerated observer in thermal bath \cite{Barman:2021oum} is
\begin{eqnarray}\label{eq:Greens-fn-TU-3D}
&& G^{\beta}_{W_{R}^{3D}}\left(\Delta\eta_{jl}\right)\nonumber\\  
&&= \int_{0}^{\infty}d\omega~\int 
\frac{ d^2\kp}{(2\pi)^4} \frac{2}{\sqrt{a_{j}a_{l}}}\nonumber\\ 
&&\left[\frac{e^{-i\omega\Delta\eta_{jl}} 
~e^{\frac{\pi\omega}{2}\left(\frac{1}{a_{j}}+\frac{1}{a_{l}}\right)}
 + e^{i\omega\Delta\eta_{jl}}~
e^{-\frac{\pi\omega}{2}\left(\frac{1}{a_{j}}+\frac{1}{a_{l}}\right)}}{1-e^{
-\beta\omega
}}\right.\nonumber\\
&& + \left.  \frac{e^{i\omega\Delta\eta_{jl}} 
~e^{\frac{\pi\omega}{2}\left(\frac{1}{a_{j}}+\frac{1}{a_{l}}\right)}
 + e^{-i\omega\Delta\eta_{jl}}~ 
e^{-\frac{\pi\omega}{2}\left(\frac{1}{a_{j}}+\frac{1}{a_{l}}\right)}}{e^{ 
\beta\omega}-1}
\right ] \nonumber\\
~&&~~~~~~\mathcal{K}\left[\frac{i\omega}{a_{j}},\frac{|\kp| 
e^{a_{j}\xi_{j}}}{a_{j}}
\right]\mathcal{K}\left[\frac{i\omega}{a_{l}},\frac{|\kp| 
e^{a_{l}\xi_{l}}}{a_{l}}
\right]~,
\end{eqnarray}
where, $\Delta\eta_{jl}=\eta_{j,2}-\eta_{l,1}$ and $\xi_{j}$ is the fixed 
Rindler spatial coordinate corresponding to the $j^{th}$ detector. It should be 
noted that the above Green's function is time translational invariant. 

On the 
other hand, the Wightman function corresponding to two observers with anti 
parallel
acceleration is
\begin{eqnarray}\label{eq:Greens-fn-TU-3D-LR}
&& G^{\beta}_{W_{LR}^{3D}}\left(\Delta\eta_{jl}\right)\nonumber\\  
&&= \int_{0}^{\infty}d\omega~\int 
\frac{ d^2\kp}{(2\pi)^4} \frac{2}{\sqrt{a_{j}a_{l}}}\nonumber\\ 
&&\left[\frac{e^{-i\omega\Delta\eta_{jl}} 
~e^{-\frac{\pi\omega}{2}\left(\frac{1}{a_{j}}-\frac{1}{a_{l}}\right)}
 + e^{i\omega\Delta\eta_{jl}}~
e^{\frac{\pi\omega}{2}\left(\frac{1}{a_{j}}-\frac{1}{a_{l}}\right)}}{1-e^{
-\beta\omega
}}\right.\nonumber\\
&& + \left.  \frac{e^{i\omega\Delta\eta_{jl}} 
~e^{-\frac{\pi\omega}{2}\left(\frac{1}{a_{j}}-\frac{1}{a_{l}}\right)}
 + e^{-i\omega\Delta\eta_{jl}}~ 
e^{\frac{\pi\omega}{2}\left(\frac{1}{a_{j}}-\frac{1}{a_{l}}\right)}}{e^{ 
\beta\omega}-1}
\right ] \nonumber\\
~&&~~~~~~\mathcal{K}\left[\frac{i\omega}{a_{j}},\frac{|\kp| 
e^{a_{j}\xi_{j}}}{a_{j}}
\right]\mathcal{K}\left[\frac{i\omega}{a_{l}},\frac{|\kp| 
e^{a_{l}\xi_{l}}}{a_{l}}
\right]~.
\end{eqnarray}
Here also $j$ and $l$ denote detectors in left and in right Rindler wedges, and 
the Wightman function $G_{W^{3D}_{RL}}^{\beta} \left(\Delta\xi_{jl}, 
\Delta\eta_{jl}\right)$, with $j$ and $l$ denoting detectors in right and left 
Rindler wedges, is obtained from the complex conjugate of the expression 
(\ref{eq:Greens-fn-TU-3D-LR}).
The thermal Green's function corresponding to a single accelerated detector can 
be obtained by making $a_{j}=a_{l}$ in Eq. (\ref{eq:Greens-fn-TU-3D}).

Having equipped with all the necessary results we will next investigate the 
role 
of temperature of the field on the entanglement harvesting between the two 
uniformly accelerated detectors. We will have particular interest here on two 
situations -- (i) both the detectors are in right wedge and (ii) one is in 
right 
wedge and another one is in left wedge. This will be done in the next section.

\section{Entanglement harvesting}\label{Entanglement-harvest}

In this section we investigate the condition of entanglement extraction from 
Eq. 
(\ref{eq:cond-entanglement}) for accelerated detectors in parallel or 
anti-parallel relative motion in a thermal bath. In particular, we aim to 
understand the effects of the thermal bath in addition to the acceleration on 
this entanglement harvesting condition. We shall also be looking into the 
entanglement measure, namely the concurrence, for the aforementioned observers. 
In this regard, we first estimate the quantities $\mathcal{I}_{j}(\Delta 
E^{j})$ 
for the detectors accelerated in right or in left Rindler wedge. These are 
common quantities for both parallel and anti-parallel situations.

We first consider the $(1+1)$ dimensional case. For an observer accelerated in 
the right Rindler wedge we take the expression of the Wightman function from 
Eq. 
(\ref{eq:Greens-fn-TU-RRW}) with equal acceleration. Then in RRW  one can 
estimate the integral $\mathcal{I}_{j}(\Delta E^{j})$ as
\begin{eqnarray}\label{eq:Ij-evaluated-RRW-1p1}
 \mathcal{I}^{R}_{j}(\Delta E^{j}) &=& \frac{1}{2} 
\int_{-\infty}^{\infty}dv_{j} 
\int_{-\infty}^{\infty}du_{j}e^{-i\Delta 
E^{j}u_{j}}~G^{\beta}_{W_{R}}(u_{j})\nonumber\\ 
~&=& \delta(0) \frac{\pi}{2\Delta E^{j}a_{j}} \frac{1}{\sinh{\frac{\pi\Delta 
E^{j}}{a_{j}}}}\Bigg[\frac{e^{-\frac{\pi\Delta 
E^{j}}{a_{j}}}}{1-e^{-\beta\Delta 
E^{j}}}\nonumber\\ 
~&&~~~~~~~~~~+~\frac{e^{\frac{\pi\Delta 
E^{j}}{a_{j}}}}{e^{\beta\Delta E^{j}}-1}\Bigg],
\end{eqnarray}
where we have used $\eta'_{j}=\tau'_{j}$, $\eta_{j}=\tau_{j}$ for 
$\xi'_{j}=0=\xi_{j}$ as observed from Eq. 
(\ref{eq:RindProper-time-acceleration}). We have considered the change of 
variables $v_{j}=\tau'_{j}+\tau_{j}$, $u_{j} = \tau'_{j} -\tau_{j}$. In the 
above expression the Dirac delta distribution is obtained from 
$\delta(0)=(1/2\pi)\int_{-\infty}^{\infty}d\gamma_{j}$, where 
$\gamma_{j}=a_{j}v_{j}$ is a dimensionless parameter characterizing the proper 
time of the detector. On the other hand, in a similar manner in LRW also one 
can 
estimate the integral $\mathcal{I}^{L}_{j}(\Delta E^{j})$ using the complex 
conjugate of the Wightman function from Eq. (\ref{eq:Greens-fn-TU-RRW}) and the 
relation between the Rindler time and the detector proper times 
$\eta'_{j}=-\tau'_{j}$, $\eta_{j}=-\tau_{j}$ in LRW for $\xi'_{j}=0=\xi_{j}$ 
from (\ref{eq:RindProper-time-acceleration}). In $(1+1)$ dimensions this 
expression comes out to be the same as the one obtained for the observer in RRW 
(\ref{eq:Ij-evaluated-RRW-1p1}), i.e., we get $\mathcal{I}^{L}_{j}(\Delta 
E^{j})=\mathcal{I}^{R}_{j}(\Delta E^{j})$.

Similarly, in $(1+3)$ dimensions also one can find out the quantities 
$\mathcal{I}^{R}_{j}(\Delta E^{j})$ and $\mathcal{I}^{L}_{j}(\Delta E^{j})$. In 
particular, these quantities in right and left Rindler wedges are given by the 
same expression $\mathcal{I}^{R}_{j}(\Delta E^{j}) = \mathcal{I}^{L}_{j}(\Delta 
E^{j}) = \mathcal{I}_{j_{3D}}(\Delta E^{j})$. With the help of the Wightman 
function from Eq. (\ref{eq:Greens-fn-TU-3D}) this expression can be provided as
\begin{eqnarray}\label{eq:Ij-evaluated-1p3}
 \mathcal{I}_{j_{3D}}(\Delta E^{j}) &=& \frac{1}{2} 
\int_{-\infty}^{\infty}dv_{j} 
\int_{-\infty}^{\infty}du_{j}~e^{-i\Delta 
E^{j}u_{j}}~G^{\beta}_{W_{3D}}(u_{j})\nonumber\\
~&=& \delta(0) \frac{1}{2\pi a_{j}^2} \Bigg[\frac{e^{-\frac{\pi\Delta 
E^{j}}{a_{j}}}}{1-e^{-\beta\Delta E^{j}}}+\frac{e^{\frac{\pi\Delta 
E^{j}}{a_{j}}}}{e^{\beta\Delta E^{j}}-1}\Bigg]\nonumber\\
~&& ~~~~~~~~~~~~~\times~ \Upsilon\left(\Delta E^{j},a_{j},a_{j}\right)~,
\end{eqnarray}
where, in this case the quantity 
$\Upsilon\left(\Delta E^{j}, a_{j},a_{j}\right)=\pi a_{j}\Delta 
E^{j}/(2\sinh{(\pi \Delta E^{j}/a_{j})})$, and it is obtained from a general 
expression of integral
\begin{equation}
\Upsilon\left(\bar{\varepsilon}, a_{j},a_{l}\right) = \int_{0}^{\infty} 
\kp~d\kp~ \mathcal{K}\left[\frac{i\bar{\varepsilon}}{a_{j}},\frac{\kp}{a_{j}} 
\right] \mathcal{K}\left[\frac{i\bar{\varepsilon}}{a_{l}},\frac{\kp}{a_{l}} 
\right].
\end{equation}
Now it should be noted that the integrals representing transition probabilities 
from Eq. (\ref{eq:Ij-evaluated-RRW-1p1}), and (\ref{eq:Ij-evaluated-1p3}) can 
be 
multiplied on both sides by $\Delta E_{j}^2$ to make them dimensionless. In 
this regard, we define other dimensionless parameters of the system as 
\begin{equation}
\alpha_{j}=\frac{a_{j}}{\Delta E_{j}}; \,\,\,\ \sigma_{j}=\beta \Delta E_{j}~.
\end{equation} 
It will be 
much more convenient to represent the necessary diagrams in our subsequent 
analysis with respect to these dimensionless parameters and quantities. In our 
subsequent analysis we specifically consider the situation of two observers 
accelerated parallelly or anti-parallelly in a thermal bath, and in particular, 
going to estimate the integrals $\mathcal{I}_{\varepsilon}$. Then we shall 
analyze the condition of Eq. (\ref{eq:cond-entanglement}), and verify the 
possibility of entanglement extraction in those  specific cases.

It should also be noted that for the verification of the condition 
(\ref{eq:cond-entanglement}) it is imperative to evaluate the expression of 
$\mathcal{I}_{\varepsilon}$. From Eq. (\ref{eq:Ie-integral}) it is observed that 
the expression of $\mathcal{I}_{\varepsilon}$ can be represented it terms of one 
quantity containing Wightman function and another one containing the Retarded 
Green's function. This second integral containing the Retarded Green's function 
also has a Heaviside step function $\theta(T_{A}-T_{B})$ in it, which emerged 
from the representation of the Feynman propagator in terms of the Wightman 
functions. In a spacetime where the Wightman functions are constructed 
considering positive frequency modes functions defined with respect to time 
$t_{j}$, this step function shall become $\theta(t_{A}-t_{B})$. From Eq. 
(\ref{eq:Greens-fn-TU-RRW}) to (\ref{eq:Greens-fn-TU-3D-LR}) all the Wightman 
functions are constructed using positive frequency Rindler modes of 
(\ref{eq:Rindler-modes}) and (\ref{eq:Rindler-modes-3D}). Then with 
$\xi'_{j}=0=\xi_{j}$ in right Rindler wedge $t_{j} = t^{\mathcal{R}}_{j} = 
\eta_{j} = \tau_{j}$ and in left Rindler wedge $t_{j} = t^{\mathcal{R}}_{j} = 
-\eta_{j} = \tau_{j}$. We shall be explicitly using these relations in the 
evaluation of the quantity $\mathcal{I}_{\varepsilon}$ in our subsequent 
analysis.

\subsection{Parallel acceleration: No harvesting}

In this subsection we consider the two observers \emph{Alice} and \emph{Bob} to 
be accelerated parallelly. We consider them to have the proper accelerations 
$a_{A}$ and $a_{B}$ and both of them to be in the right Rindler wedge. For 
convenience of calculation, in this case we express integral 
$\mathcal{I}_{\varepsilon}$ from Eq. (\ref{eq:Ie-integral}) as
\begin{eqnarray}\label{eq:Ie-integral-2}
 && \mathcal{I}_{\varepsilon} = -\int_{-\infty}^{\infty}d\tau_{B} 
\int_{-\infty}^{\infty}d\tau_{A}~\scalebox{0.91}{$e^{i(\Delta 
E^{B}\tau_{B}+\Delta E^{A}\tau_{A})} $}\times\nonumber\\
~&& \scalebox{0.87}{$\left[G_{W}(X_{B},X_{A})+ 
\theta(t^{\mathcal{R}}_{A}-t^{\mathcal{R}}_{B}) 
\left\{G_{W}\left(X_{A},X_{B}\right)-G_{W}\left(X_{B},X_{A}\right)\right\}
\right]$}\nonumber\\
~&& ~~=~\mathcal{I}^{W}_{\varepsilon}~+~\mathcal{I}^{R}_{\varepsilon}~.
\end{eqnarray}
Here the first integral $\mathcal{I}^{W}_{\varepsilon}$ contains the Wightman 
function, while the second integral $\mathcal{I}^{R}_{\varepsilon}$ represents 
the contribution of the retarded Green's function. We shall be using this 
expression to evaluate the integral $\mathcal{I}_{\varepsilon}$ separately in 
$(1+1)$ and $(1+3)$ dimensions in our following studies.

\subsubsection{$(1+1)$ dimensions}

For the evaluation of $\mathcal{I}^{W}_{\varepsilon}$ and 
$\mathcal{I}^{R}_{\varepsilon}$ in $(1+1)$ dimensions we consider the positive 
frequency Wightman function (\ref{eq:Greens-fn-TU-RRW}). In particular, in the 
expression of this Green's function the indices $j$ and $l$ correspond to the 
detector $A$ and $B$ respectively. The relation between Rindler times and 
detector proper times are $\eta_{A}=\tau_{A}$ and $\eta_{B}=\tau_{B}$, 
considering $\xi_{A}=0=\xi_{B}$, i.e., assuming the accelerating detectors to 
be 
fixed at the origin of the respective Rindler frames, while the proper 
accelerations are $b_{j}=a_{j}$. Then the first integral 
$\mathcal{I}^{W}_{\varepsilon}$ can be expressed as
\begin{eqnarray}\label{eq:IeW-parallel-1p1}
 && \mathcal{I}^{W}_{\varepsilon} = -\int_{-\infty}^{\infty}d\tau_{B} 
\int_{-\infty}^{\infty}d\tau_{A}~\scalebox{0.91}{$e^{i(\Delta 
E^{B}\tau_{B}+\Delta E^{A}\tau_{A})} $}\times\nonumber\\
&& ~~~~~~~~~~~~~~~~~~~~~~~~~~G^{\beta}_{W_{R}}(X_{B},X_{A})\nonumber\\
&=&-\delta\left(\frac{\Delta E^A+\Delta 
E^B}{\sqrt{a_Aa_B}}\right) 
\frac{\pi}{\Delta\breve{E}\sqrt{a_Aa_B}}\frac{1}{\sqrt{\sinh{\frac{
\pi\Delta\breve {E}}{ a_B}}\sinh{\frac{\pi\Delta\breve{E}}{a_A}}} 
}\nonumber\\
&&~~~~~~~~~~\left[\frac{e^{\frac{\pi\Delta\breve{E}}{2}\left(\frac{1 } 
{a_{B}}+\frac{1}{a_{A}}\right)}}{1-e^{-\beta\Delta\breve{E} 
}}+\frac{e^{-\frac{\pi\Delta\breve{E}}{2}\left(\frac{1}{a_{B}}+\frac{1}{a_{A}} 
\right)}}{e^{\beta\Delta\breve{E}}-1}\right]~,
\end{eqnarray}
where the expression of $\Delta\breve{E}$ is given by $\Delta\breve{E} = (\Delta 
E^{B} - \Delta E^{A})/2$. For the evaluation of this integral we have considered 
a change of variables $\tilde{v}=\tau_{B}+\tau_{A}$ and 
$\tilde{u}=\tau_{B}-\tau_{A}$. The Jacobian of this transformation from 
$\tau_{j}$ to $v$ and $u$ is $1/2$. On the other hand, using the same Wightman 
function from Eq. (\ref{eq:Greens-fn-TU-RRW}) we get the  integral 
$\mathcal{I}^{R}_{\varepsilon}$ to be
\begin{eqnarray}\label{eq:IeR-parallel-1p1}
 \mathcal{I}^{R}_{\varepsilon} &=& -\int_{-\infty}^{\infty}d\tau_{B} 
\int_{-\infty}^{\infty}d\tau_{A}~\scalebox{0.91}{$e^{i(\Delta 
E^{B}\tau_{B}+\Delta E^{A}\tau_{A})} $}\times\nonumber\\
&&~~\theta(\tau_{A}-\tau_{B})\left\{G^{\beta}_{W_{R}}(X_{A},X_{B})-G^{
\beta}_{W_{R}} (X_ { B } , X_ { A } 
)\right\}\nonumber\\
 &=&-\delta\left(\frac{\Delta E^A+\Delta 
E^B}{\sqrt{a_Aa_B}}\right) \mathscr{I}^{R}_{1},
\end{eqnarray}
which also contains a Dirac delta distribution, and $\mathscr{I}^{R}_{1}$ is 
given by
\begin{eqnarray}
 \mathscr{I}^{R}_{1} &=& \int_{0}^{\infty}
\frac{d\omega}{\omega\sqrt{a_Aa_B}} \frac{\sinh\left[ 
\frac{\pi\omega}{2}\left(\frac{1}{a_B}+\frac{1}{a_A}\right)\right]}{ 
\sqrt 
{\sinh{\frac{\pi\omega}{a_B}}\sinh{\frac{\pi\omega}{a_A}}}}\nonumber\\
~&&~~~~~~~~\times
 \int_{0}^{\infty} e^{-i\tilde{u}\Delta 
\breve{E}} \left(e^{-i\omega\tilde{u}}-e^{i\omega\tilde{u}}\right)~.
\end{eqnarray}
In the integrals (\ref{eq:IeW-parallel-1p1}) and (\ref{eq:IeR-parallel-1p1}) we 
have encountered expression $\delta((\Delta E^{B}+\Delta E^{A})/\sqrt{a_{A} 
a_{B}})$, which definitely cannot give non-zero contribution as $\Delta 
E^{j}>0$. Then we have $\mathcal{I}_{\varepsilon}=0$ due to the contributions of 
$\mathcal{I}^{W}_{\varepsilon}$ and $\mathcal{I}^{R}_{\varepsilon}$. Whereas 
from Eq. (\ref{eq:Ij-evaluated-RRW-1p1}) we observe that $\mathcal{I}_{j}$ 
always are non-zero and also have a multiplicative $\delta(0)$. This 
signifies that for two observers accelerated parallelly the condition for 
entanglement harvesting (\ref{eq:cond-entanglement}) is not satisfied in $(1+1)$ 
dimensions.

\subsubsection{$(1+3)$ dimensions}

We consider the positive frequency Wightman function (\ref{eq:Greens-fn-TU-3D}) 
for the estimation of the quantities $\mathcal{I}^{W}_{\varepsilon}$ and 
$\mathcal{I}^{R}_{\varepsilon}$ in $(1+3)$ dimensions. Furthermore, we have 
identified the indices $j$ and $l$ with the detectors $A$ and $B$ respectively. 
Then proceeding like the earlier way the first integral 
$\mathcal{I}^{W}_{\varepsilon}$ becomes 
\begin{eqnarray}\label{eq:IeW-parallel-1p3}
 && \mathcal{I}^{W}_{\varepsilon} = -\int_{-\infty}^{\infty}d\tau_{B} 
\int_{-\infty}^{\infty}d\tau_{A}~\scalebox{0.91}{$e^{i(\Delta 
E^{B}\tau_{B}+\Delta E^{A}\tau_{A})} $}\nonumber\\
&& ~~~~~~~~~~~~~~~~~~~~~~~~~~G^{\beta}_{W^{3D}_{R}}(X_{B},X_{A})\nonumber\\
&=&-\delta\left(\frac{\Delta E^A+\Delta 
E^B}{\sqrt{a_Aa_B}}\right) 
\frac{1}{\pi a_A a_B} 
\Upsilon\left(\Delta\breve{E},a_{A},a_{B}\right)\nonumber\\
&&~~~~~~\left[\frac{e^{\frac{\pi\Delta\breve{E}}{2}\left(\frac{1} 
{a_{B}}+\frac{1}{a_{A}}\right)}}{1-e^{-\beta\Delta\breve{E}
}}+\frac{e^{-\frac{\pi\Delta\breve{E}}{2}\left(\frac{1}{a_{B}}+\frac{1}{a_{A}}
\right)}}{e^{\beta\Delta\breve{E}}-1}\right].
\end{eqnarray}
Using the Wightman function (\ref{eq:Greens-fn-TU-3D}) with the proper
identification of the indices $j$ and $l$ to the detectors $A$ and $B$, 
we get the second part of the integral $\mathcal{I}_{\varepsilon}$ to be 
\begin{eqnarray}\label{eq:IeR-parallel-1p3}
 \mathcal{I}^{R}_{\varepsilon} &=& -\int_{-\infty}^{\infty}d\tau_{B} 
\int_{-\infty}^{\infty}d\tau_{A}~\scalebox{0.91}{$e^{i(\Delta 
E^{B}\tau_{B}+\Delta E^{A}\tau_{A})} $}\times\nonumber\\
&& 
~~\theta(\tau_{A}-\tau_{B})\left\{G^{\beta}_{W^{3D}_{R}}(X_{A},X_{B})-G^
{\beta}_{W^{3D}_{R}}(X_{B},X_{
A})\right\}\nonumber\\
&=& -\delta\left(\frac{\Delta E^{B}+\Delta 
E^{A}}{\sqrt{a_Aa_B}}\right)  \mathscr{I}^{R}_{2}
~,
\end{eqnarray}
where, the expression of $ \mathscr{I}^{R}_{2}$is given by 
\begin{eqnarray}
 \mathscr{I}^{R}_{2} &=& \int_{0}^{\infty}
\frac{d\omega}{\pi^2} \frac{\sinh\left[ 
\frac{\pi\omega}{2}\left(\frac{1}{a_B}+\frac{1}{a_A}\right)\right]}{a_A a_B
}\Upsilon\left(\Delta\breve{E},a_{A},a_{B}\right)\nonumber\\
~&&~~~~~~~~\times
 \int_{0}^{\infty} e^{-i\tilde{u}\Delta 
\breve{E}} \left(e^{-i\omega\tilde{u}}-e^{i\omega\tilde{u}}\right)~.
\end{eqnarray}
Here also in both of the integrals from (\ref{eq:IeW-parallel-1p3}) and 
(\ref{eq:IeR-parallel-1p3}) we have $\delta((\Delta E^{B}+\Delta 
E^{A})/\sqrt{a_{A} a_{B}})$ multiplied, which always gives zero contribution 
when $\Delta E^{j}>0$. Then none of the above quantities 
$\mathcal{I}^{W}_{\varepsilon}$ or $\mathcal{I}^{R}_{\varepsilon}$ contribute 
to 
the non-zero expression of $\mathcal{I}_{\varepsilon}$. This signifies that for 
two observers accelerated parallelly, the condition for entanglement harvesting 
is not satisfied in $(1+3)$ dimensions.

In both $(1+1)$ and $(1+3)$ dimensions, considering parallelly accelerated 
detectors in a thermal bath, we observed that the condition for entanglement 
harvesting is not satisfied. This was also true in the case of accelerated 
detectors without a thermal bath, see \cite{Reznik:2002fz, Koga:2018the, 
Koga:2019fqh}. Then one can deduce that here {\it the thermal bath has no 
additional influence to make the entanglement harvesting possible}. One should 
also notice that, it is not possible to distinguish between thermal and 
non-thermal scalar fields only by analyzing the parallelly accelerated 
detectors 
using entanglement harvesting information.

\subsection{Anti-parallel accelerations: harvesting possible}

In this subsection, we consider \emph{Alice} in right and \emph{Bob} in the left 
Rindler wedge so that they are anti-parallelly accelerated in a thermal bath. To 
evaluate the integral $\mathcal{I}_{\varepsilon}$ and to reflect upon the 
previously obtained result by Reznik \cite{Reznik:2002fz}, where only the 
Wightman function rather than the Feynman propagator contributed to this 
integral, it is convenient to express it in the form of Eq. 
(\ref{eq:Ie-integral-2}).
%
%
Here also we shall be separately evaluating the first integral 
$\mathcal{I}^{W}_{\varepsilon}$ and the second integral 
$\mathcal{I}^{R}_{\varepsilon}$, which respectively contains the Wightman 
function and the retarded Green's function,  in $(1+1)$ and $(1+3)$ dimensions.

\subsubsection{(1+1)dimensions}
\underline{\it Analytical results:} -- 
In $(1+1)$ dimensions the first part $\mathcal{I}^{W}_{\varepsilon}$ of the 
integral $\mathcal{I}_{\varepsilon}$, is estimated using the expression of the 
Wightman function from Eq. (\ref{eq:Greens-fn-TU-LR}). We have also considered 
\emph{Bob} (denoted by $B$) to be in LRW accelerating anti-parallelly to 
\emph{Alice} in RRW (denoted by $A$). Then the integral 
$\mathcal{I}^{W}_{\varepsilon}$ becomes
\begin{eqnarray}\label{eq:IE-1p1-1}
\mathcal{I}^{W}_{\varepsilon} &=& -\int_{-\infty}^{\infty}d\tau_{B} 
\int_{-\infty}^{\infty}d\tau_{A}~\scalebox{0.91}{$e^{i(\Delta 
E^{B}\tau_{B}+\Delta E^{A}\tau_{A})} G^{\beta}_{W_{LR}}(X_{B},X_{A})
$}\nonumber\\
&=&-\delta\left(\frac{\Delta E^{B}-\Delta 
E^{A}}{\sqrt{a_{A}a_{B}}}\right)\frac{1}{\sqrt{\sinh{\frac{\pi\Delta 
\widetilde{E}}{a_{A}}}\sinh{\frac{\pi\Delta 
\widetilde{E}}{a_{B}}}}}\nonumber\\
~&& \frac{\pi}{\Delta 
\widetilde{E}\sqrt{a_{A}a_{B}}}\Bigg[\frac{e^{\frac{\pi\Delta 
\widetilde{E}}{2}\left(\frac{1}{a_{B}}-\frac{1}{a_{A}}\right)}}{1-e^{
-\beta\Delta 
\widetilde{E}}}+\frac{e^{-\frac{\pi\Delta 
\widetilde{E}}{2}\left(\frac{1}{a_{B}}-\frac{1}{a_{A}}\right)}}{e^{\beta\Delta 
\widetilde{E}}-1}\Bigg]~,\nonumber\\
\end{eqnarray}
where, $\Delta \widetilde{E}=(\Delta E^{B}+\Delta E^{A})/2$, and $\delta(z)$ 
denotes the Dirac delta distribution. For the evaluation of this integral we 
have made the  change of variables $\tilde{v}=\tau_{B}+\tau_{A}$ and 
$\tilde{u}= 
\tau_{B} -\tau_{A}$. One may have  considered moving to dimensionless variables 
$\gamma_{j}=\tau_{j}a_{j}$, and then make change of variables 
$\bar{v}=\gamma_{B}+\gamma_{A}$ and $\bar{u}=\gamma_{B}-\gamma_{A}$ to obtain 
the same final result. The Jacobian corresponding to both of these change of 
variables is $1/2$. Similarly one can evaluate the second integral 
$\mathcal{I}^{R}_{\varepsilon}$. Then using the Wightman function of Eq. 
(\ref{eq:Ie-integral}) the integral $\mathcal{I}^{R}_{\varepsilon}$ can be 
evaluated to be
\begin{eqnarray}\label{eq:IE-1p1-2a}
\mathcal{I}^{R}_{\varepsilon} &=& -\int_{-\infty}^{\infty}d\tau_{B} 
\int_{-\infty}^{\infty}d\tau_{A}\theta(\tau_{A}-\tau_{B})~e^{
i(\Delta 
E^{B}\tau_{B}+\Delta E^{A}\tau_{A})} \nonumber\\
~&& ~~~~~~~~~~~\big[  G^{\beta}_{W_{RL}}(X_{A},X_{B})- 
G^{\beta}_{W_{LR}}(X_{B},X_{A}) \big]\nonumber\\
&=&\frac{\sinh{\frac{\pi\Delta 
\widetilde{E}}{2}\left(\frac{1}{a_{B}}-\frac{1}{a_{A}}\right)}}{2\Delta 
\widetilde{E}}\frac{1}{\sqrt{\sinh{\frac{\pi\Delta 
\widetilde{E}}{a_{A}}}\sinh{\frac{\pi\Delta 
\widetilde{E}}{a_{B}}}}}\nonumber\\
~&&~~~~~~~~~ \times\int_{-\infty}^{\infty}e^{-\frac{i}{2}(\Delta E^{B}-\Delta 
E^{A})\tilde{u}}~\theta(\tilde{u})~d\tilde{u}~,
\end{eqnarray}
where, $\Delta \widetilde{E}=(\Delta E_{A} + \Delta E_{B})/2$. One can evaluate 
this last integral as 
\begin{eqnarray}\label{eq:IE-1p1-2b}
 \int_{-\infty}^{\infty}\scalebox{0.9}{$e^{-\frac{i}{2}(\Delta E^{B}-\Delta 
E^{A})\tilde{u}}~\theta(\tilde{u})d\tilde{u} $} &=& \int_{0}^{\infty} 
\scalebox{0.9}{$e^{-\frac{i}{2}(\Delta E^{B}-\Delta 
E^{A}-i\epsilon)\tilde{u}}~d\tilde{u}$}\nonumber\\
~&=& \scalebox{1}{$\frac{2}{i(\Delta E^{B}-\Delta 
E^{A}-i\epsilon)}~,$}
\end{eqnarray}
where a multiplicative regulator of $e^{-\epsilon \tilde{u}/2}$, with 
$\epsilon>0$, is introduced in the integrand to evaluate this otherwise 
diverging integral. It is to be noted that the limit $\epsilon\to0$ provides 
the 
actual value of the integral. One can express this quantity of Eq. 
(\ref{eq:IE-1p1-2b}) with the help of a consequence of the 
\emph{Sokhotski-Plemelj theorem} \cite{book:Birrell}
\begin{equation}\label{eq:IE-1p1-2c}
 \lim_{\epsilon\to0+}\frac{1}{z-i\epsilon} = i\pi~ 
\delta(z)+\mathcal{P}\left(\frac{1}{z}\right) ~,
\end{equation}
where, $\mathcal{P}(1/z)$ denotes the principal value of $(1/z)$, which is a 
finite quantity. Then in the limit of $\Delta E^{B}\to\Delta E^{A}$ the only 
contributing quantity in $\mathcal{I}^{R}_{\varepsilon}$ is
\begin{eqnarray}\label{eq:IE-1p1-2d}
\mathcal{I}^{R}_{\varepsilon} &=&\frac{\pi\sinh{\left\{\frac{\pi\Delta 
\widetilde{E}}{2}\left(\frac{1}{a_{B}}-\frac{1}{a_{A}}\right)\right\}}}{\Delta 
\widetilde{E}\sqrt{a_{A}a_{B}}\sqrt{\sinh{\frac{\pi\Delta 
\widetilde{E}}{a_{A}}}\sinh{\frac{\pi\Delta 
\widetilde{E}}{a_{B}}}}}\nonumber\\
~&& \times~\Big[\delta\left(\tfrac{\Delta 
E^{B}-\Delta E^{A}}{\sqrt{a_{A}a_{B}}}\right)-\frac{i}{\pi} 
\mathcal{P}\left(\tfrac{\sqrt{a_{A}a_{B}}}{\Delta 
E^{B}-\Delta E^{A}}\right)\Big].
\end{eqnarray}
The second quantity of the multiplicative term in the right hand side of this 
expression denotes the principle value and it is a finite quantity. It should 
be 
mentioned that when $\Delta E^{B}\neq \Delta E^{A}$ the Dirac delta 
distribution 
from Eq. (\ref{eq:IE-1p1-2d}) vanishes and one is left out with only this 
finite 
second term. In this particular situation the integral 
$\mathcal{I}^{W}_{\varepsilon}$ also vanishes and the whole 
$\mathcal{I}_{\varepsilon}= \mathcal{I}^{W}_{\varepsilon} + 
\mathcal{I}^{R}_{\varepsilon}$ becomes finite. However, from Eq. 
(\ref{eq:Ij-evaluated-RRW-1p1}) it is observed that the integrals 
$\mathcal{I}_{j}$ still keeps the $\delta(0)$ terms in them. In that case it is 
obvious that the condition from (\ref{eq:cond-entanglement}) remains 
unfulfilled. On the other hand, when $\Delta E^{B}= \Delta E^{A}$ only the 
Dirac 
delta distribution contributes in the expression of  
$\mathcal{I}^{R}_{\varepsilon}$. In this situation, 
$\mathcal{I}^{W}_{\varepsilon}$ is also non zero, and there are multiplicative 
factors of $\delta(0)$ in $\mathcal{I}^{W}_{\varepsilon}$ and 
$\mathcal{I}^{R}_{\varepsilon}$ like the $\mathcal{I}_{j}$. Then it is evident 
that only for $\Delta E^{B}= \Delta E^{A}$ the condition 
(\ref{eq:cond-entanglement}) for entanglement harvesting may get satisfied.

%
%

Let us now make a comment on the contribution related to the retarded Green's 
function. It is observed from (\ref{eq:IE-1p1-2d}) that the integral 
$\mathcal{I}^{R}_{\varepsilon}$ in general vanishes when the two anti 
parallelly 
moving observers have equal magnitude of accelerations i.e. $a_{A}=a_{B}$. It 
is 
noticed that even in Green's function level when the accelerations of the 
anti-parallelly accelerated detectors are equal the quantity 
$G^{\beta}_{W_{RL}}(X_{A},X_{B})- G^{\beta}_{W_{LR}}(X_{B},X_{A}) = 0$, 
denoting 
the retarded Green's function in the integral of (\ref{eq:IE-1p1-2a}). This is 
expected as left Rindler wedge is causally disconnected from the right wedge. 
However, it remains non-zero for scenarios when $a_{A}\neq a_{B}$, although 
both 
are causally disconnected and retarded Green's function with respect to 
Minkowski mode vanishes (or when $a_A=0=a_B$) when they are spacelike 
separated. 
It may be mentioned that when $a_A=a_B$ then they can be regarded as ``similar 
frames'' (as LRW is mirror image of RRW) and hence since retarded Green's 
function vanishes in Minkowski frame, it must vanish in any other frame. 
Whereas 
for $a_A\neq a_B$ we do not have such similarity and we may take this as 
investigations of field operators from two ``dis-similar frames''. In this case 
the commutator of the fields may not be the same as it was earlier. In the 
above exactly this thing has been reflected in $\mathcal{I}^{R}_{\varepsilon}$. 
In summary, the relative acceleration between the frames introduces this 
non-triviality. We will see later that in $(1+3)$ case, compared to $(1+1)$ 
dimensional analysis, this has a big role to give a distinct feature in the 
entanglement harvesting. 

  From Eq. 
(\ref{eq:IE-1p1-1}) and (\ref{eq:IE-1p1-2d}) we obtain the expression of the 
integral $\mathcal{I}_{\varepsilon}$ corresponding to two anti-parallelly 
accelerated observers as $\mathcal{I}_{\varepsilon}= 
\mathcal{I}^{W}_{\varepsilon} +\mathcal{I}^{R}_{\varepsilon}$. As we have 
already discussed these expressions are non-zero and comparable to 
$\mathcal{I}_{j}$ only when $\Delta E^{A}$ and $\Delta E^{B}$ are equal, we 
then 
consider $\Delta E^{A}=\Delta E^{B}=\Delta E$. In that case we also have 
$\Delta 
\widetilde{E}=\Delta E$, and the condition for entanglement 
harvesting from Eq. (\ref{eq:cond-entanglement}) becomes
\begin{eqnarray}\label{eq:Cond-EntHarvest-ab-1p1}
&& \scalebox{1.1}{$\Bigg(\frac{e^{-\frac{\pi\Delta 
E}{a_{A}}}}{1-e^{-\beta\Delta E}}+\frac{e^{\frac{\pi\Delta 
E}{a_{A}}}}{e^{\beta\Delta E}-1}\Bigg)  \Bigg(\frac{e^{-\frac{\pi\Delta 
E}{a_{B}}}}{1-e^{-\beta\Delta E}}+\frac{e^{\frac{\pi\Delta 
E}{a_{B}}}}{e^{\beta\Delta E}-1}\Bigg)$}<\nonumber\\
~&& ~~~~~~~~~~~~4\scalebox{1.1}{$\Bigg[\frac{e^{\frac{\pi\Delta 
E}{2}\left(\frac{1}{a_{B}}-\frac{1}{a_{A}}\right)}}{1-e^{-\beta\Delta 
E}}+\frac{e^{-\frac{\pi\Delta 
E}{2}\left(\frac{1}{a_{B}}-\frac{1}{a_{A}}\right)}}{e^{\beta\Delta 
E}-1}$}\nonumber\\
~&& ~~~~~~~~~~~~~~~~\scalebox{1}{$-\sinh{\left\{\frac{\pi\Delta 
E}{2}\left(\frac{1}{a_{B}}-\frac{1}{a_{A}}\right)\right\}}\Bigg]^2$}~.
\end{eqnarray}
%
\begin{figure}[h]
\centering
 \includegraphics[width=1.05\linewidth]{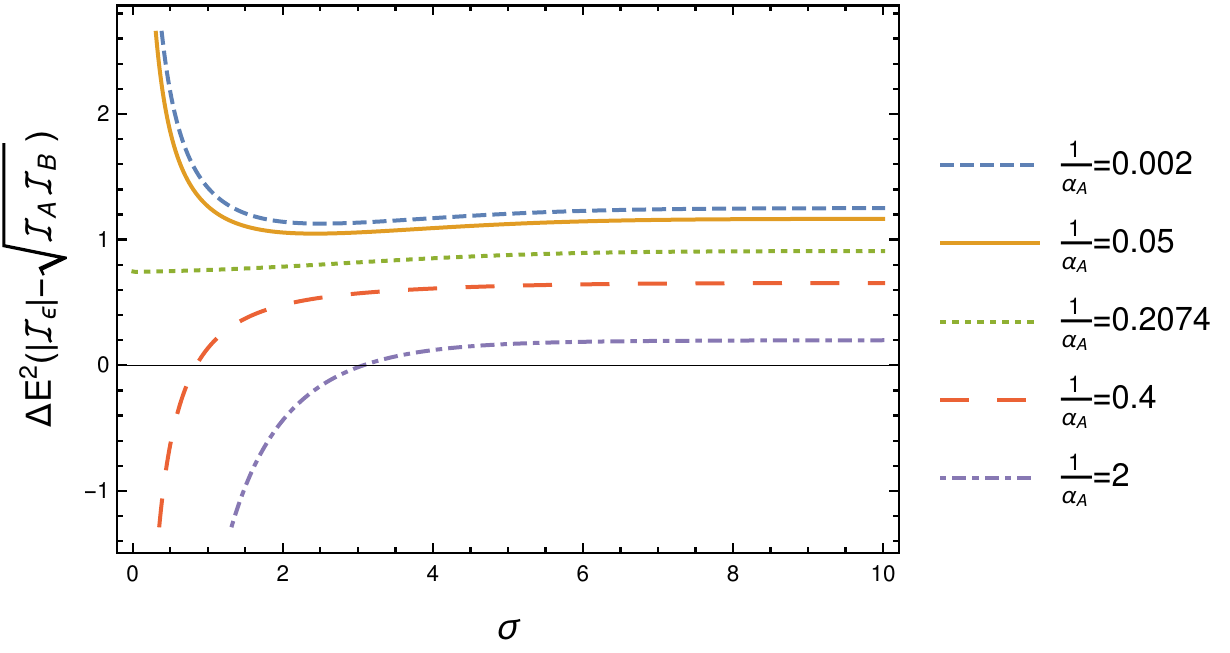}
 \caption{In $(1+1)$ dimensions the quantity $\Delta 
E^{2}\left(|\mathcal{I}_{\varepsilon}|- \sqrt{\mathcal{I}_{A} 
\mathcal{I}_{B}}\right)$ is plotted for two anti-parallelly accelerating 
detectors with respect to the inverse temperature $\sigma=\beta\Delta E$ for 
different fixed  $\alpha_{A}=a_{A}/\Delta E$. The other parameter is fixed at 
$\alpha_{B}=a_{B}/\Delta E=1$.}
 \label{fig:EHC-1p1-ab1-Vbeta}
\end{figure}
%
From this expression (\ref{eq:Cond-EntHarvest-ab-1p1}) depicting the condition 
for entanglement harvesting for two anti-parallelly accelerated observers, we 
see that contribution of the retarded Green's function exists when the 
detectors 
have different magnitudes of acceleration, i.e., $a_{A}\neq a_{B}$. 

\underline{\it Numerical analysis:} --
In Fig.  \ref{fig:EHC-1p1-ab1-Vbeta} we have plotted the quantity 
$\mathcal{C}_{\mathcal{I}}$ of (\ref{eq:concurrence-I}) signifying the 
concurrence, with respect to $\sigma$, which is proportional to the inverse 
temperature of the thermal bath $\sigma=\beta\Delta E$, considering two 
anti-parallelly accelerated observers with different accelerations in $(1+1)$ 
dimensions. The curves in this figure correspond to fixed $\alpha_{B}=1$ and 
different fixed $\alpha_{A}$. It is to be noted that the quantity 
$\mathcal{C}_{\mathcal{I}}$ plotted in the figure is obtained using the 
expressions from (\ref{eq:Ij-evaluated-RRW-1p1}), (\ref{eq:IE-1p1-1}) and 
(\ref{eq:IE-1p1-2d}) when $\Delta E^{B}= \Delta E^{A}$ without the 
multiplicative $\delta(0)$ term. Removing this delta function from 
$\mathcal{C}_{\mathcal{I}}$ can be interpreted as taking a rate per unit proper 
time, like discussed in literature \cite{Koga:2019fqh}. From Fig.  
\ref{fig:EHC-1p1-ab1-Vbeta} we have the following observations.

\begin{itemize}
 \item For low acceleration $\alpha_{A}$ of the first detector (e.g. 
$\alpha_{A}=1/2$) the quantity $\mathcal{C}_{\mathcal{I}}$ is negative for very 
high temperature of the thermal bath, and it tends to increase with increasing 
$\beta$ and becomes positive at some much large $\beta$ or low temperature of 
the thermal bath. Therefore thermal fields do not allow entanglement at high 
temperature. Entanglement can start only from certain value of temperature of 
thermal bath to lower values when the first detector moves with small 
acceleration. 

 \item For high acceleration $\alpha_{A}$ of the first detector 
($\alpha_{A}=1/0.002$) the quantity $\mathcal{C}_{\mathcal{I}}$ is positive for 
very high temperature of the thermal bath, and it tends to decrease with 
increasing $\beta$ but never becomes negative at much larger $\beta$ or low 
temperature of the thermal bath. So for large values of acceleration, we will 
have entanglement at any temperature of bath.
 
 \item With these it is observed that there is a characteristic change in the 
nature of these curves depending on the value of $\alpha_{A}$ -- for low values 
of $\alpha_A$ the entanglement increases with increase of $\beta$ while after 
certain value of $\alpha_{A}$ entanglement decreases with increase of $\beta$. 
We call the value of acceleration $\alpha_{A}=\alpha_{c}$ as critical value 
around  which these curves have different nature. In Fig.  
\ref{fig:EHC-1p1-ab1-Vbeta} this is given by the almost straight line which is 
green in color.

\end{itemize}
To find this critical value note that the change is nature is prominent for 
very 
low value of $\beta$; i.e. at higher temperature of the bath. So it will be 
sufficient to investigate $\mathcal{C}_{\mathcal{I}}$ for very low value of 
$\sigma$. Also in this regime the critical line (which is green in Fig.   
\ref{fig:EHC-1p1-ab1-Vbeta}) is straight. Therefore the vanishing of the slope 
of the curve in low value of $\sigma$ will yield $\alpha_c$. One can series 
expand the derivative of $\mathcal{C}_{\mathcal{I}}$ with respect to $\sigma$ 
in 
this case, in small $\sigma$ regimes and observe that it is of the form 
$(\partial/\partial\sigma)\mathcal{C}(\rho_{AB})= (1/\sigma^2) \mathcal{C}_{1} 
+ 
\mathcal{C}_{2} + \mathcal{O}(\sigma)$. Then in high temperature regime one can 
predict about the transition point $\alpha_{c}$ by making $\mathcal{C}_{1} = 0$ 
(this is leading term) with $\alpha_A=\alpha_c$, which provides us with the 
expression
\begin{eqnarray}
 && \mathcal{C}_{1} = \frac{\pi}{ \sqrt{\sinh \left(\frac{\pi 
}{\alpha_{c}}\right) \sinh \left(\frac{\pi 
}{\alpha_{B}}\right)}} \times\nonumber\\
~&& \scalebox{0.95}{$   
\left(\sqrt{\cosh \left(\frac{\pi }{\alpha_{c}}\right) \cosh \left(\frac{\pi 
}{\alpha_{B}}\right)}-2 \cosh \left(\frac{\pi  (\alpha_{B}-\alpha_{c})}{2 
\alpha_{c} \alpha_{B}}\right)\right) $}=0~.\nonumber\\
\end{eqnarray}
It can be checked that the above equation yields the value of $\alpha_c$ as 
$4.82026$ for our choice of parameter value $\alpha_B=1$. Note that this is 
exactly the value of $\alpha_A$ for which the critical curve (green in color in 
Fig. \ref{fig:EHC-1p1-ab1-Vbeta}) was obtained numerically.\vspace{0.2cm}
\begin{figure}[h]
\centering
 \includegraphics[width=1.02\linewidth]{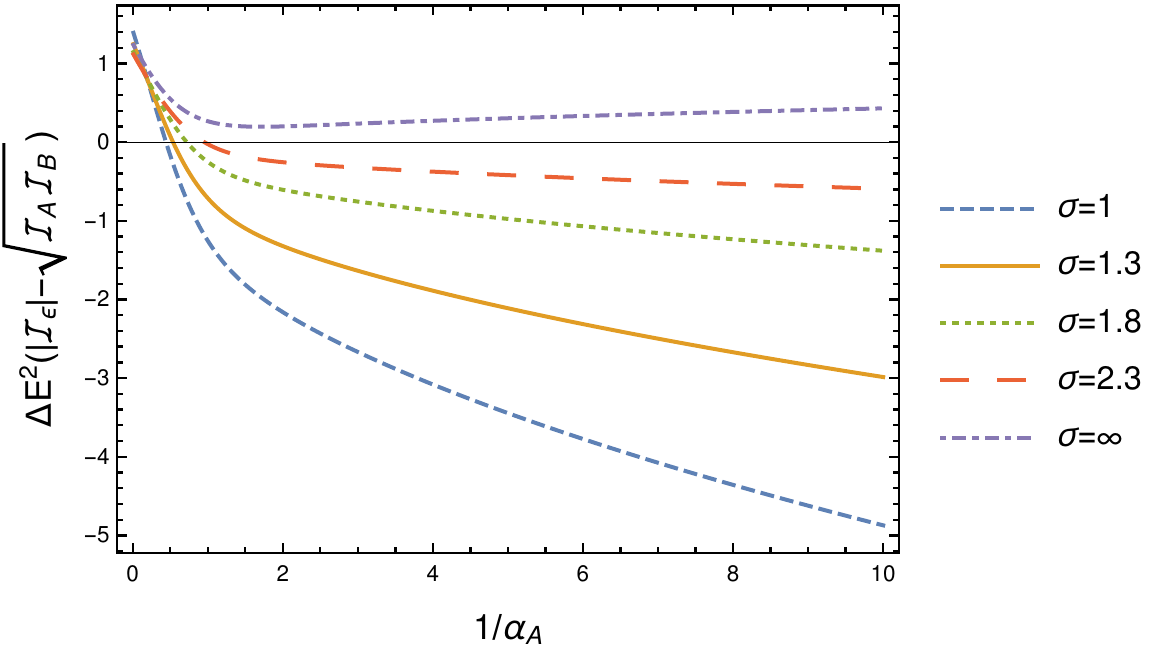}
 \caption{In $(1+1)$ dimensions the quantity $\Delta 
E^{2}\left(|\mathcal{I}_{\varepsilon}|- \sqrt{\mathcal{I}_{A} 
\mathcal{I}_{B}}\right)$ is plotted for two anti-parallelly accelerating 
detectors with respect to the acceleration of the first detector $\alpha_{A}$ 
for different fixed inverse temperature of the thermal bath $\sigma$. The other 
parameters are fixed at $\alpha_{B}=1$.}
\label{fig:EHC-1p1-ab1-Va}
\end{figure}

\begin{figure}[h]
\centering
 \includegraphics[width=1.05\linewidth]{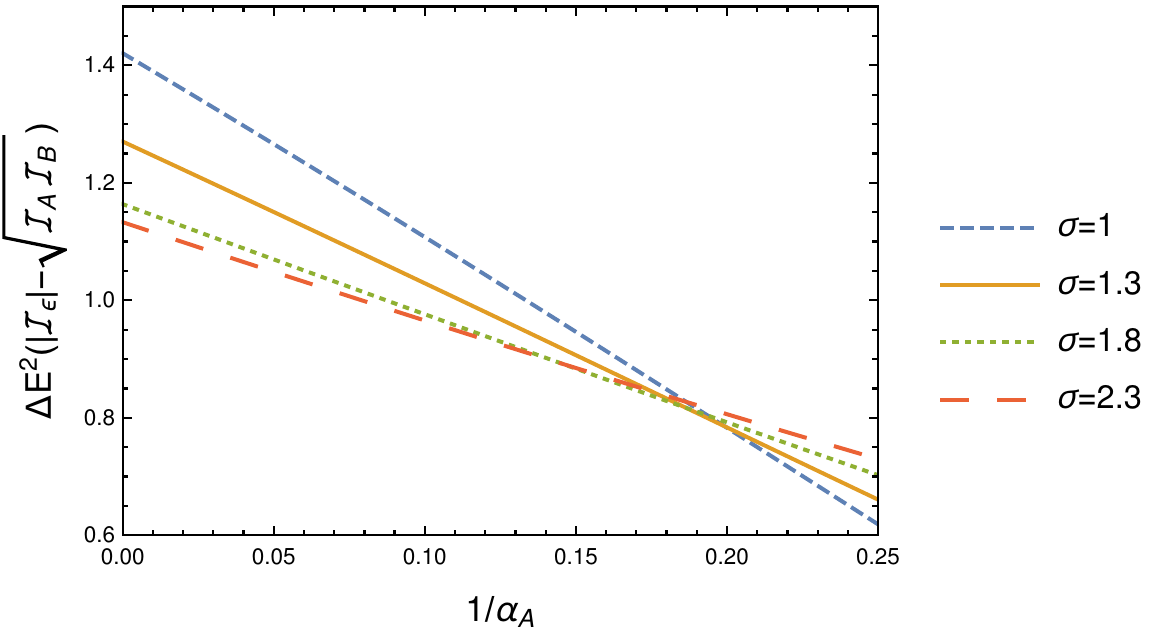}
 \caption{In $(1+1)$ dimensions the quantity $\Delta 
E^{2}\left(|\mathcal{I}_{\varepsilon}|- \sqrt{\mathcal{I}_{A} 
\mathcal{I}_{B}}\right)$ is plotted for two anti-parallelly accelerating 
detectors with respect to the acceleration of the first detector $\alpha_{A}$ 
for different fixed inverse temperatures of the thermal bath $\sigma$. The 
other 
parameters are fixed at $\alpha_{B}=1$. In particular we have depicted the 
curves of Fig. \ref{fig:EHC-1p1-ab1-Va} in lower regime of $1/\alpha_{A}$.}
\label{fig:EHC-1p1-ab1-Va2}
\end{figure}

\begin{figure}[h]
\centering
 \includegraphics[width=1.05\linewidth]{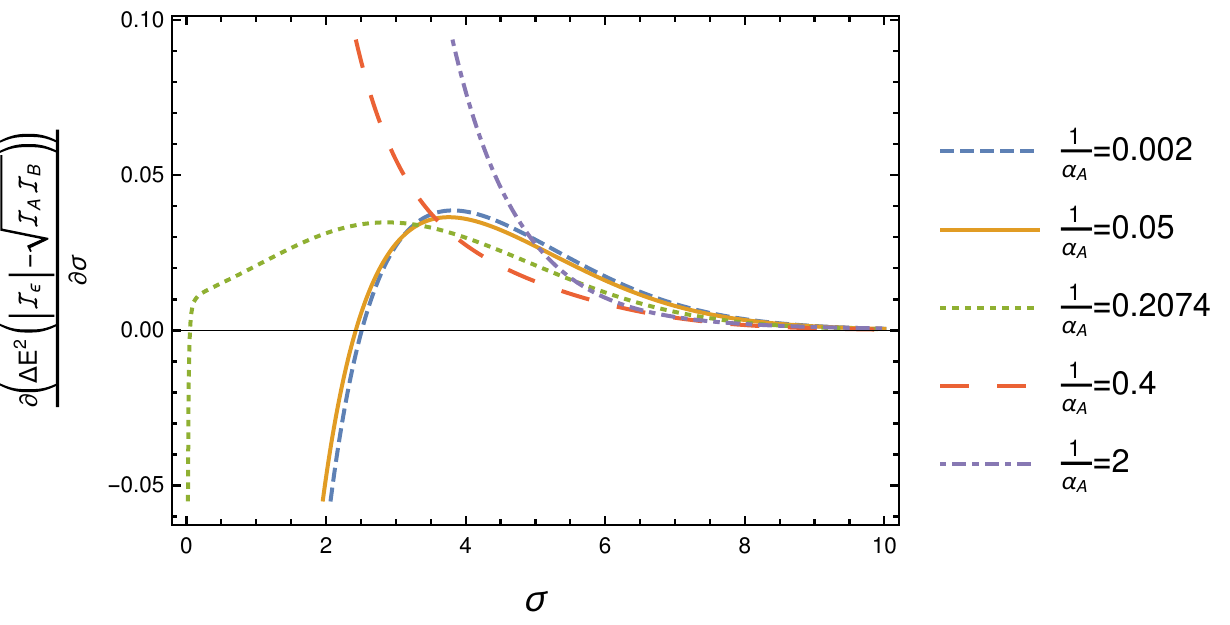}
 \caption{In $(1+1)$ dimensions the derivative with respect to $\sigma$ of the 
quantity $\Delta E^{2}\left(|\mathcal{I}_{\varepsilon}|- \sqrt{\mathcal{I}_{A} 
\mathcal{I}_{B}}\right)$ is plotted for two anti-parallelly accelerating 
detectors for varying $\sigma$. The other parameters $\alpha_{B}=1$ and 
$\alpha_{A}$ are fixed.}
\label{fig:DEHC-1p1-ab1-Vbeta}
\end{figure}

\begin{figure}[h]
\centering
 \includegraphics[width=1.05\linewidth]{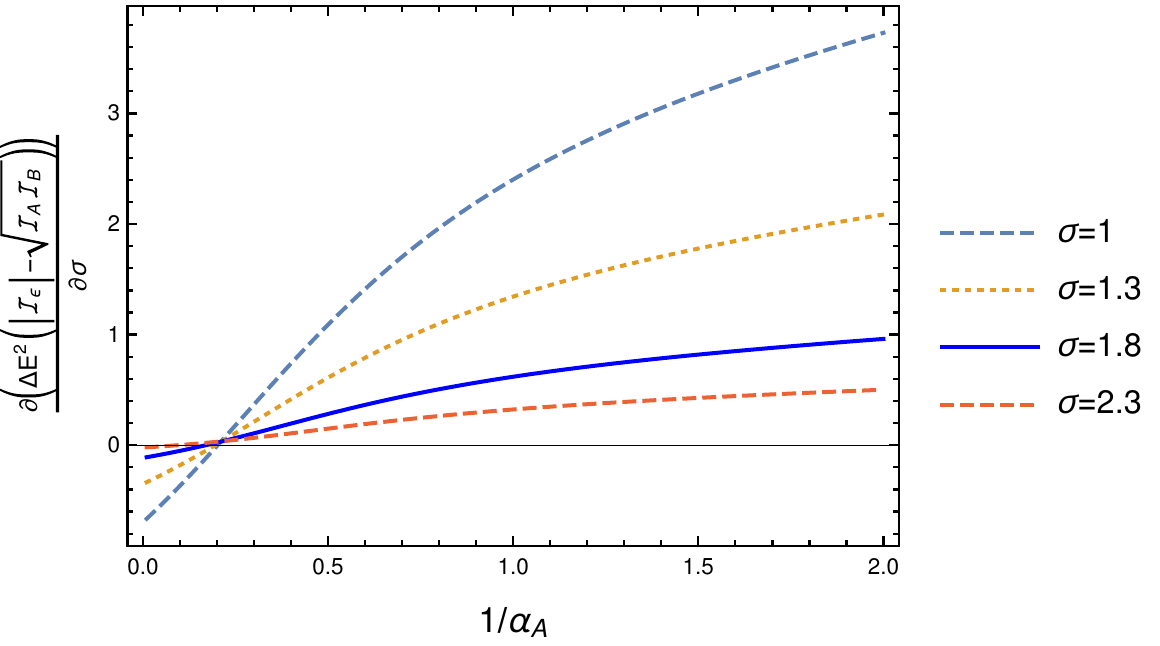}
 \caption{In $(1+1)$ dimensions the derivative with respect to $\sigma$ of the 
quantity $\Delta E^{2}\left(|\mathcal{I}_{\varepsilon}|- \sqrt{\mathcal{I}_{A} 
\mathcal{I}_{B}}\right)$ is plotted for two anti-parallelly accelerating 
detectors for varying acceleration of the first detector $\alpha_{A}$. The 
temperatures of the thermal bath $\sigma$ and other parameter $\alpha_{B}=1$ 
are 
fixed.}
\label{fig:DEHC-1p1-ab-Va}
\end{figure}

In Fig. \ref{fig:EHC-1p1-ab1-Va} we have plotted $\mathcal{C}_{\mathcal{I}}$ 
signifying the concurrence with respect to the acceleration of the first 
detector $\alpha_{A}$ for different fixed $\sigma$. From this figure one can 
observe that the temperature of the thermal bath has a diminishing effect on the 
entanglement measure for low values of the acceleration of the first detector 
$\alpha_{A}$. It is observed that for smaller $\alpha_{A}$ with low $\beta$ (if 
$\Delta E$ is kept fixed then $\beta$ changes in unison with $\sigma$), i.e., 
for very high temperature, the condition for entanglement harvesting is failing, 
while for high $\alpha_{A}$ the condition again gets satisfied. We have also 
depicted the same curves as shown in Fig. \ref{fig:EHC-1p1-ab1-Va} in lower 
regimes of $1/\alpha_{A}$ in Fig. \ref{fig:EHC-1p1-ab1-Va2}. From this curve we 
observe that, above a certain value of $\alpha_{A}$, which is the critical value 
$\alpha_{c}$, the thermal bath has an enhancing effect on concurrence (denoted 
by $\mathcal{C}_{\mathcal{I}}$). Then the plots depicted in Fig. 
\ref{fig:EHC-1p1-ab1-Va} and \ref{fig:EHC-1p1-ab1-Va2} together predict the same 
phenomena provided by Fig. \ref{fig:EHC-1p1-ab1-Vbeta}, i.e., for low 
accelerations thermal bath has a diminishing effect and for high accelerations 
thermal bath has an enhancing effect on the entanglement measure, and there is a 
perceivable critical value of acceleration separating these two regimes of 
accelerations. In Fig. \ref{fig:DEHC-1p1-ab1-Vbeta} and \ref{fig:DEHC-1p1-ab-Va} 
we have further plotted the derivative of $\mathcal{C}_{\mathcal{I}}$ with 
respect to $\sigma$ for varying $\sigma$ and $\alpha_{A}$ for the perception of 
$\alpha_{c}$. Fig. \ref{fig:DEHC-1p1-ab1-Vbeta} shows that some curves contains 
negative slope while others have positive slope for initial values of $\beta$. 
Similarly Fig. \ref{fig:DEHC-1p1-ab-Va} signifies that the derivative of the 
quantity, denoting concurrence, with respect to $\sigma$ becomes zero  at a 
particular value of $\alpha_A$. All these reassured the existence the aforesaid 
critical value of $\alpha_A$. \vspace{0.2cm}

\begin{figure}[h]
\centering
 \includegraphics[width=1.02\linewidth]{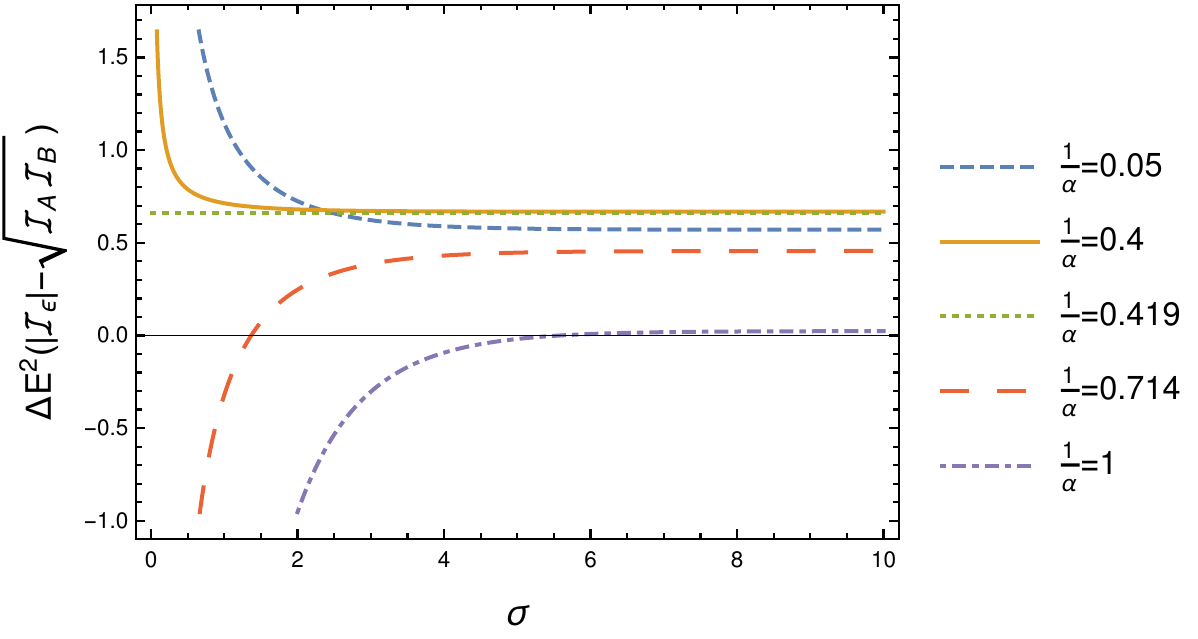}
 \caption{In $(1+1)$ dimensions the quantity $\Delta 
E^2\left(|\mathcal{I}_{\varepsilon}|- \sqrt{\mathcal{I}_{A} 
\mathcal{I}_{B}}\right)$ is plotted for two anti-parallelly accelerating 
detectors with respect to the inverse temperature of the thermal bath $\sigma$ 
for equal magnitude of proper accelerations, i.e., $\alpha_{A}=\alpha_{B}$.} 
\label{fig:EHC-1p1-aa-Vbeta}
\end{figure}

\begin{figure}[h]
\centering
 \includegraphics[width=1.05\linewidth]{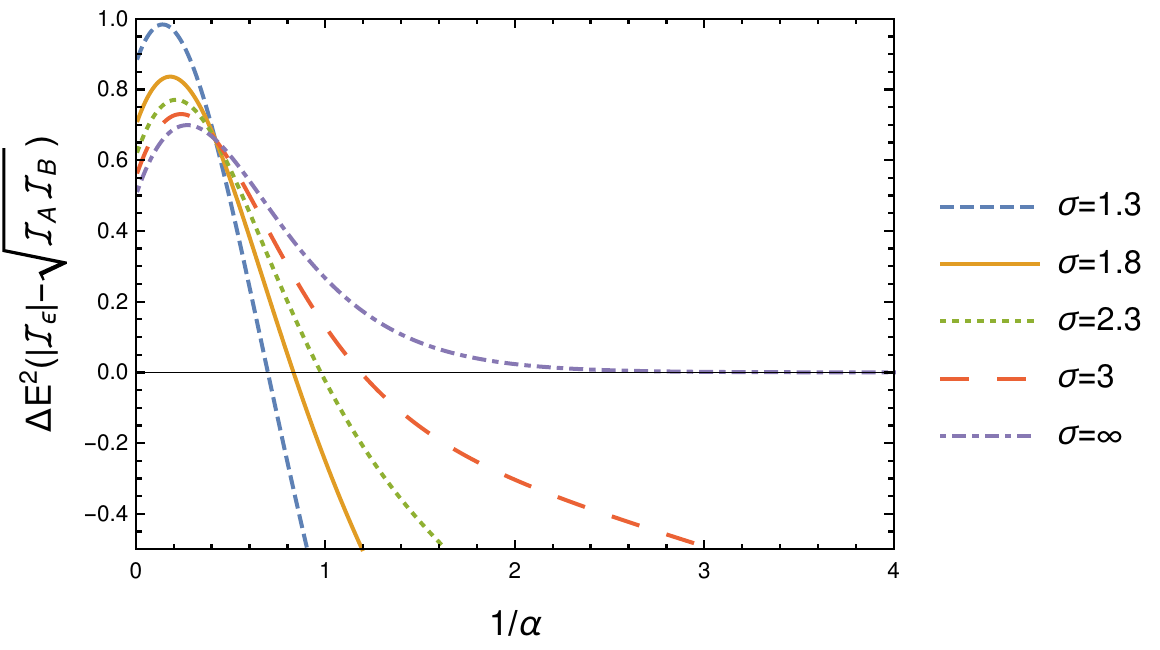}
 \caption{In $(1+1)$ dimensions the quantity $\Delta 
E^2\left(|\mathcal{I}_{\varepsilon}|- \sqrt{\mathcal{I}_{A} 
\mathcal{I}_{B}}\right)$ is plotted for two anti-parallelly accelerating 
detectors with same magnitude of acceleration for varying acceleration of the 
detectors $\alpha$ and different fixed $\sigma$.}
\label{fig:EHC-1p1-aa-Va}
\end{figure}

\begin{figure}[h]
\centering
 \includegraphics[width=1.05\linewidth]{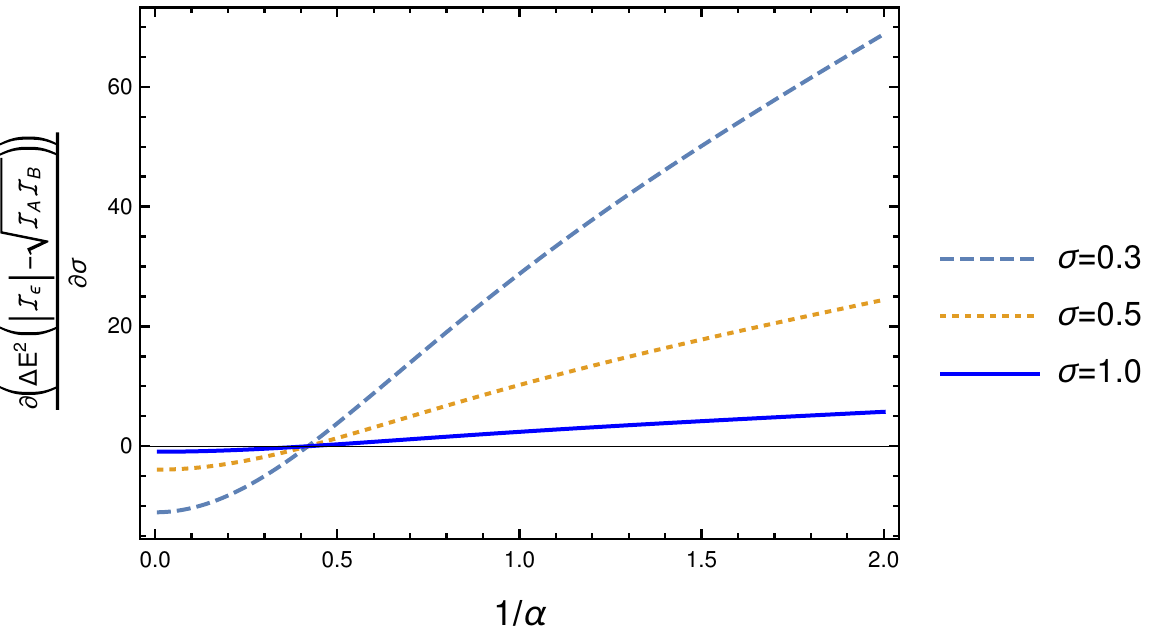}
 \caption{In $(1+1)$ dimensions the quantity $(\partial/\partial\sigma) 
\left(\Delta E^2\left(|\mathcal{I}_{\varepsilon}|- \sqrt{\mathcal{I}_{A} 
\mathcal{I}_{B}}\right)\right)$ is plotted for two anti-parallelly accelerating 
detectors with respect to the equal magnitude of proper accelerations 
$\alpha_{A}=\alpha_{B}$ for fixed inverse temperature of the thermal bath 
$\sigma$. The critical acceleration, where this quantity is $\sigma$ 
independent, is $\alpha_{c}=2.3854$.}
\label{fig:DEHC-1p1-Va}
\end{figure}

\begin{figure}[h]
\centering
 \includegraphics[width=1.05\linewidth]{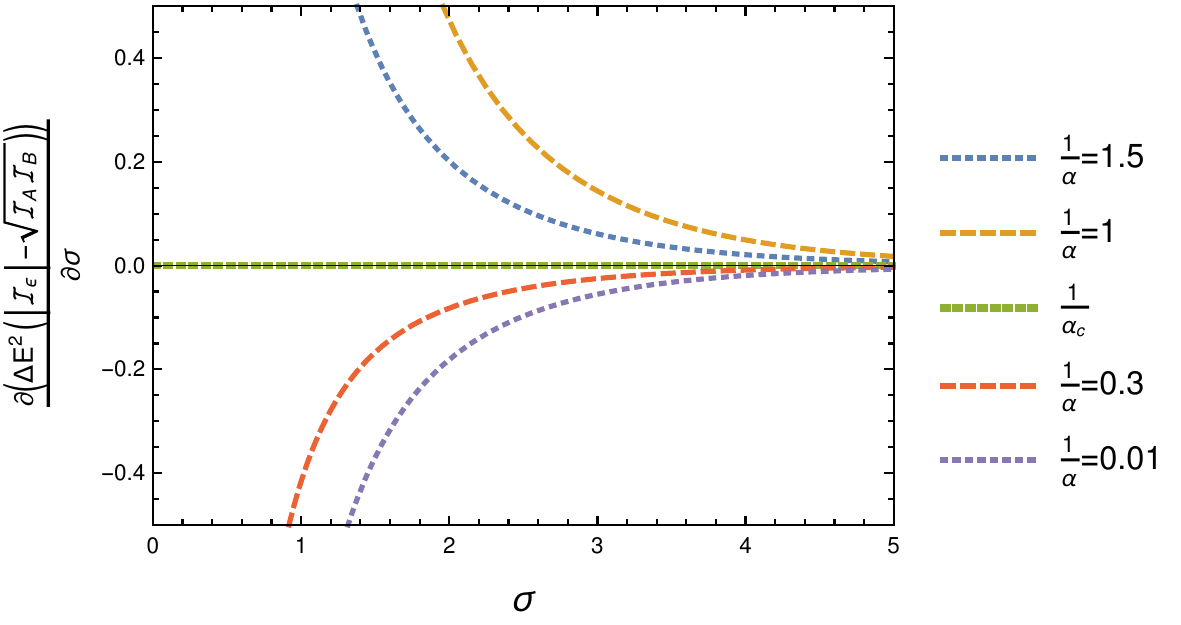}
 \caption{In $(1+1)$ dimensions the quantity $(\partial/\partial\sigma) 
\left(\Delta E^2\left(|\mathcal{I}_{\varepsilon}|- \sqrt{\mathcal{I}_{A} 
\mathcal{I}_{B}}\right)\right)$ is plotted for two anti-parallelly accelerating 
detectors with respect to the inverse temperature of the thermal bath $\sigma$ 
for equal magnitude of proper accelerations $\alpha_{A}=\alpha_{B}$. The 
critical acceleration, where this quantity is $\sigma$ independent, is 
$\alpha_{c}=2.3854$.}
\label{fig:DEHC-1p1-Vbeta}
\end{figure}

It is to be noted that in the equal magnitude of acceleration limit the second 
integral from Eq. (\ref{eq:IE-1p1-2d}) coming from the retarded Green's 
function 
vanishes and one is left with only $\mathcal{I}_{\varepsilon}= 
\mathcal{I}^{W}_{\varepsilon}$. In this particular case we 
consider $a_{A}=a_{B}=a$, and the condition for entanglement harvesting from 
Eq. (\ref{eq:cond-entanglement}) is then given by 
\begin{eqnarray}\label{eq:Cond-EntHarvest-aa-1p1}
\frac{e^{-\frac{\pi\Delta E}{a}}}{1-e^{-\beta\Delta 
E}}+\frac{e^{\frac{\pi\Delta E}{a}}}{e^{\beta\Delta 
E}-1}<2~ \frac{e^{\beta\Delta 
E}+1}{e^{\beta\Delta E}-1}~,
\end{eqnarray}
which for the zero temperature of the thermal bath, i.e., in the 
$\beta\to\infty$ limit, becomes $ e^{\frac{\pi\Delta E}{a}}>1/2$. This 
basically 
reinstates the fact that in the zero temperature case the entanglement can be 
harvested for anti-parallelly accelerated detectors with any possible equal 
acceleration, which is also observed from \cite{Reznik:2002fz, Koga:2019fqh} 
though considering the detectors in $(1+3)$ dimensions. In Fig. 
\ref{fig:EHC-1p1-aa-Vbeta} and Fig. \ref{fig:EHC-1p1-aa-Va} we have plotted 
$\mathcal{C}_{\mathcal{I}}=  |\mathcal{I}_{\varepsilon}| -\mathcal{I}_{j}$ 
respectively with respect to varying $\sigma$ and $\alpha$. From these figures 
also we observe the same phenomena as perceived before. Here also we see that 
below a certain critical $\alpha$ entanglement harvesting is not possible for 
low $\beta$ or high temperature of the thermal bath and the entanglement 
measure 
increases with increasing $\beta$. On the other hand, above this critical 
acceleration entanglement measure decreases with increasing $\beta$, but 
remains 
positive. In Fig. \ref{fig:EHC-1p1-aa-Va} this behavioral change of the curves 
after a certain critical acceleration $\alpha_{c}$ is much more prominent than 
the previous ones with different accelerations. It should be noted that in this 
equal acceleration case, by making the derivative of the quantity 
$\mathcal{C}_{\mathcal{I}}$ with respect to $\sigma$ equal to zero, one can 
obtain the critical value of acceleration $\alpha_{c}= \pi/ 
\log{[2+\sqrt{3}]}$, 
which is around $\alpha_{c}\approx2.385$ and is independent of $\sigma$. This 
is 
depicted by a straight line in Fig. \ref{fig:EHC-1p1-aa-Vbeta}. In Fig. 
\ref{fig:DEHC-1p1-Va} and \ref{fig:DEHC-1p1-Vbeta} the derivative of 
$\mathcal{C}_{\mathcal{I}}$ is plotted with respect to varying $\alpha$ and 
$\sigma$, which also signifies the earlier mentioned slope change about the 
critical value of $\alpha$. This reconfirms the existence of the aforesaid 
criticality.

\subsubsection{(1+3)dimensions}
\underline{\it Analytical results:} --
From Eq. (\ref{eq:Ie-integral}) we observe that there are two specific terms in 
the integral $\mathcal{I}_{\varepsilon}$. One involving a Wightman function and 
another involving a retarded Green's function. The second term involving the 
retarded Green's function is recently conceived through rigorous analysis of 
the 
model for entanglement harvesting. Like the $(1+1)$ dimensional case in $(1+3)$ 
dimensions also we shall explicitly evaluate these terms. We express the first 
quantity using the Wightman function of Eq. (\ref{eq:Greens-fn-TU-3D-LR}), 
i.e., 
considering the observer $B$ to be accelerating anti-parallelly in LRW with 
respect to observer $A$ in RRW, as
\begin{eqnarray}\label{eq:IE-1p3-1}
\mathcal{I}^{W}_{\varepsilon} &=& -\int_{-\infty}^{\infty}d\tau_{B} 
\int_{-\infty}^{\infty}d\tau_{A}~\scalebox{0.91}{$e^{i(\Delta 
E^{B}\tau_{B}+\Delta E^{A}\tau_{A})} G^{\beta}_{W^{3D}_{LR}}(X_{B},X_{A})
$}\nonumber\\
&=&-\delta\left(\frac{\Delta E^{B}-\Delta 
E^{A}}{\sqrt{a_{A}a_{B}}}\right)\frac{1}{a_{A}a_{B}}
\frac{ \Upsilon\left(\Delta \widetilde{E}, 
a_{B},a_{A}\right)}{\pi}\nonumber\\
~&& ~~\times\Bigg[\frac{e^{\frac{\pi\Delta 
\widetilde{E}}{2}\left(\frac{1}{a_{B}}-\frac{1}{a_{A}}\right)}}{1-e^{
-\beta\Delta 
\widetilde{E}}}+\frac{e^{-\frac{\pi\Delta 
\widetilde{E}}{2}\left(\frac{1}{a_{B}}-\frac{1}{a_{A}}\right)}}{e^{\beta\Delta 
\widetilde{E}}-1}\Bigg]~,
\end{eqnarray}
where, $\Delta \widetilde{E}=(\Delta E^{B}+\Delta E^{A})/2$, and $\delta(z)$ 
denotes the Dirac delta distribution. For the evaluation of this integral we 
have considered change of variables $\tilde{v}=\tau_{B}+\tau_{A}$ and 
$\tilde{u}=\tau_{B}-\tau_{A}$, and we shall be using this same change of 
variables to evaluate the next integral also. Then one can evaluate the second 
part of the integral $\mathcal{I}_{\varepsilon}$ from Eq. 
(\ref{eq:Ie-integral}) 
as
\begin{eqnarray}\label{eq:IE-1p3-2a}
\mathcal{I}^{R}_{\varepsilon} &=& -\int_{-\infty}^{\infty}d\tau_{B} 
\int_{-\infty}^{\infty}d\tau_{A}\theta(\tau_{A}-\tau_{B})~e^{
i(\Delta 
E^{B}\tau_{B}+\Delta E^{A}\tau_{A})} \nonumber\\
~&& ~~~~~~~~~~\big[ 
G^{\beta}_{W^{3D}_{RL}}(X_{A},X_{B})-G^{\beta}_{W^{3D}_{LR}}(X_{B},X_{A}) 
\big]\nonumber\\
&=&\sinh{\left\{\frac{\pi\Delta 
\widetilde{E}}{2}\left(\frac{1}{a_{B}}-\frac{1}{a_{A}}\right)\right\}}~ \frac{2 
\Upsilon\left(\Delta \widetilde{E}, 
a_{B},a_{A}\right)}{(2\pi)^2\sqrt{a_{A}a_{B}}}\nonumber\\
~&&~~~~~~~~~ \times\int_{-\infty}^{\infty}e^{-\frac{i}{2}(\Delta E^{B}-\Delta 
E^{A})\tilde{u}}~\theta(\tilde{u})~d\tilde{u}~.
\end{eqnarray}
Like the previous $(1+1)$ dimensional case, using (\ref{eq:IE-1p1-2c}) the 
contributing part of this expression here can be evaluated to be
\begin{eqnarray}\label{eq:IE-1p3-2d}
\mathcal{I}^{R}_{\varepsilon} &=&
\frac{\sinh{\left\{\frac{\pi\Delta 
\widetilde{E}}{2}\left(\frac{1}{a_{B}}-\frac{1}{a_{A}}\right)\right\}}
\Upsilon\left(\Delta 
\widetilde{E}, a_{B},a_{A}\right)}{\pi a_{A}a_{B}}\nonumber\\
~&& \times\Big[\delta\left(\tfrac{\Delta 
E^{B}-\Delta E^{A}}{\sqrt{a_{A}a_{B}}}\right)-\frac{i}{\pi} 
\mathcal{P}\left(\tfrac{\sqrt{a_{A}a_{B}}}{\Delta 
E^{B}-\Delta E^{A}}\right)\Big].
\end{eqnarray}
%
\begin{figure}[h]
\centering
 \includegraphics[width=1.05\linewidth]{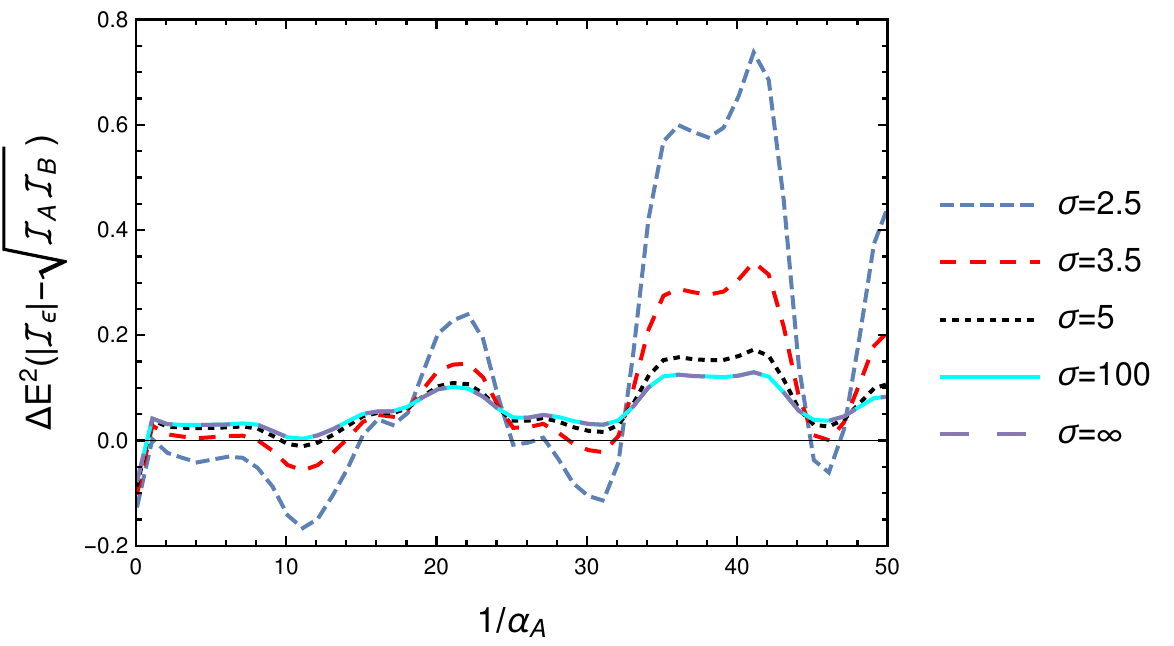}
 \caption{In $(1+3)$ dimensions the quantity $\Delta 
E^2\left(|\mathcal{I}_{\varepsilon}|- \sqrt{\mathcal{I}_{A} 
\mathcal{I}_{B}}\right)$ is plotted for two anti-parallelly accelerating 
detectors with respect to the acceleration of the first detector $\alpha_{A}$ 
for different fixed $\sigma$. The other parameter is fixed at $\alpha_{B}=1$.}
 \label{fig:EHC-1p3-ab1-Va}
\end{figure}
\begin{figure}[h]
\centering
 \includegraphics[width=1.07\linewidth]{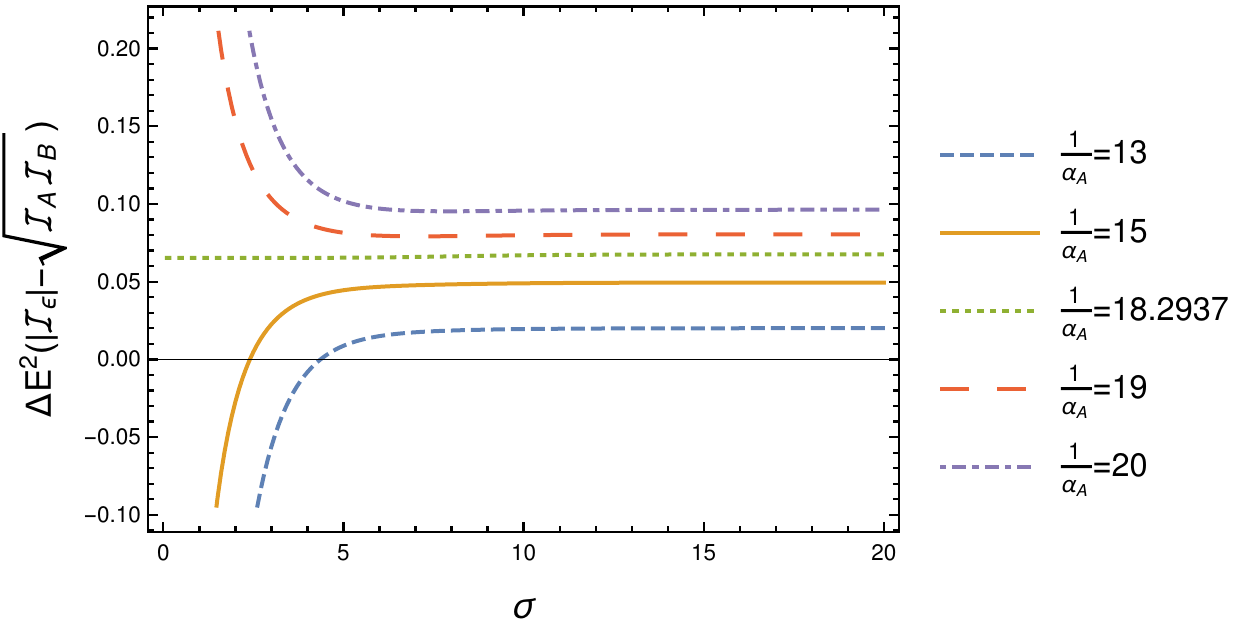}
 \caption{In $(1+3)$ dimensions the quantity $\Delta 
E^2\left(|\mathcal{I}_{\varepsilon}|- \sqrt{\mathcal{I}_{A} 
\mathcal{I}_{B}}\right)$ is plotted for two anti-parallelly accelerating 
detectors with respect to the inverse temperature of the thermal bath $\sigma$ 
for different fixed accelerations $\alpha_{A}$. The other parameter is fixed at 
$\alpha_{B}=1$.}
 \label{fig:EHC-1p3-ab1-Vbeta}
\end{figure}
\begin{figure}[h]
\centering
 \includegraphics[width=1.05\linewidth]{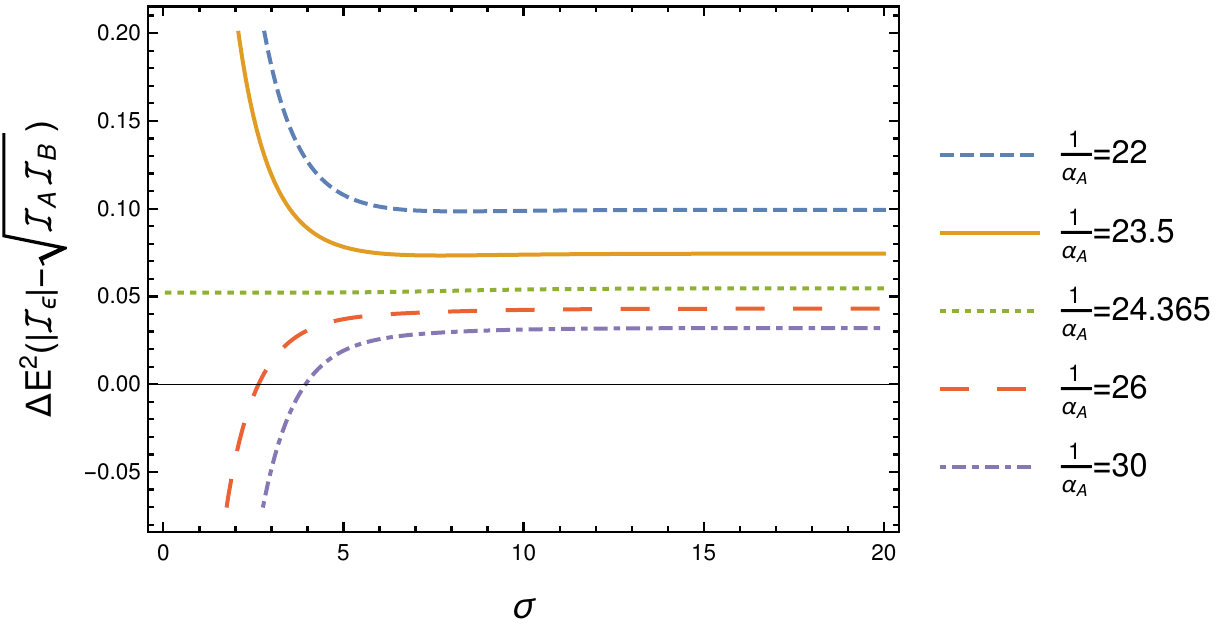}
 \caption{In $(1+3)$ dimensions the quantity $\Delta 
E^2\left(|\mathcal{I}_{\varepsilon}|- \sqrt{\mathcal{I}_{A} 
\mathcal{I}_{B}}\right)$ is plotted for two anti-parallelly accelerating 
detectors with respect to the inverse temperature of the thermal bath $\sigma$ 
for different fixed accelerations $\alpha_{A}$, and $\alpha_{B}$ is fixed 
at $\alpha_{B}=1$. Here the set of fixed $\alpha_{A}$ is different than the 
ones considered in Fig. \ref{fig:EHC-1p3-ab1-Vbeta}. However, here also one can 
observe a transition in the nature of the curves as $\alpha_{A}$ changes.}
 \label{fig:EHC-1p3-ab1-Vbeta2}
\end{figure}

\begin{figure}[h]
\centering
 \includegraphics[width=1.05\linewidth]{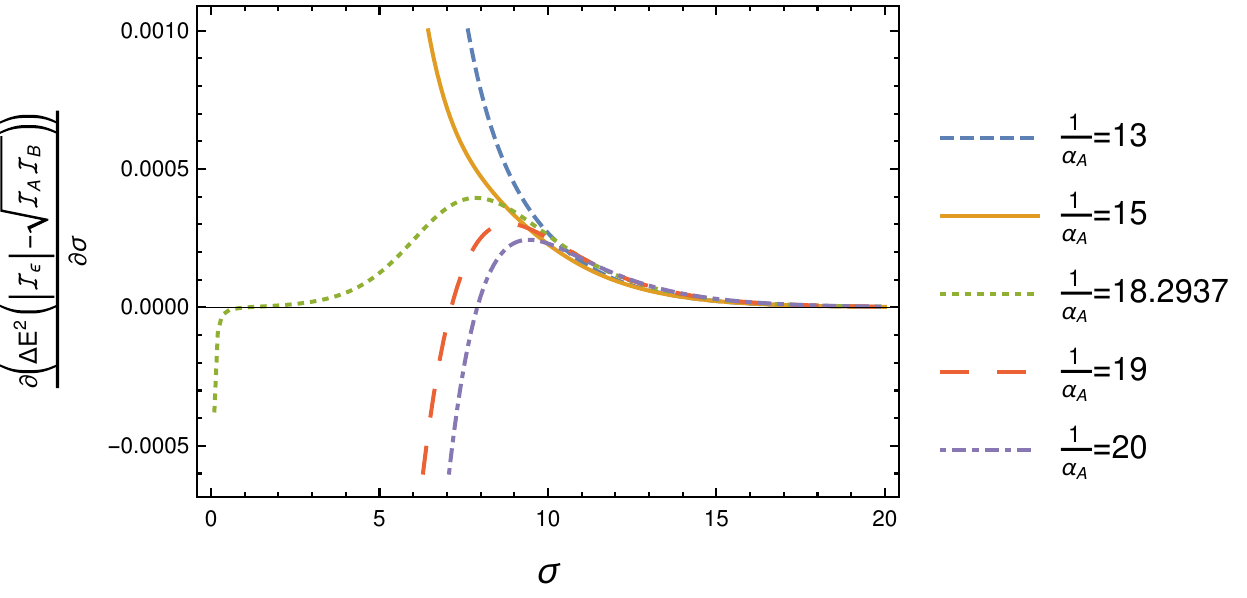}
 \caption{In $(1+3)$ dimensions the derivative with respect to $\sigma$ of the 
quantity $\Delta E^{2}\left(|\mathcal{I}_{\varepsilon}|- \sqrt{\mathcal{I}_{A} 
\mathcal{I}_{B}}\right)$ is plotted for two anti-parallelly accelerating 
detectors for varying $\sigma$. The other parameters $\alpha_{B}=1$ and 
$\alpha_{A}$ are fixed.}
\label{fig:DEHC-1p3-ab-Vbeta}
\end{figure}

\begin{figure}[h]
\centering
 \includegraphics[width=1.05\linewidth]{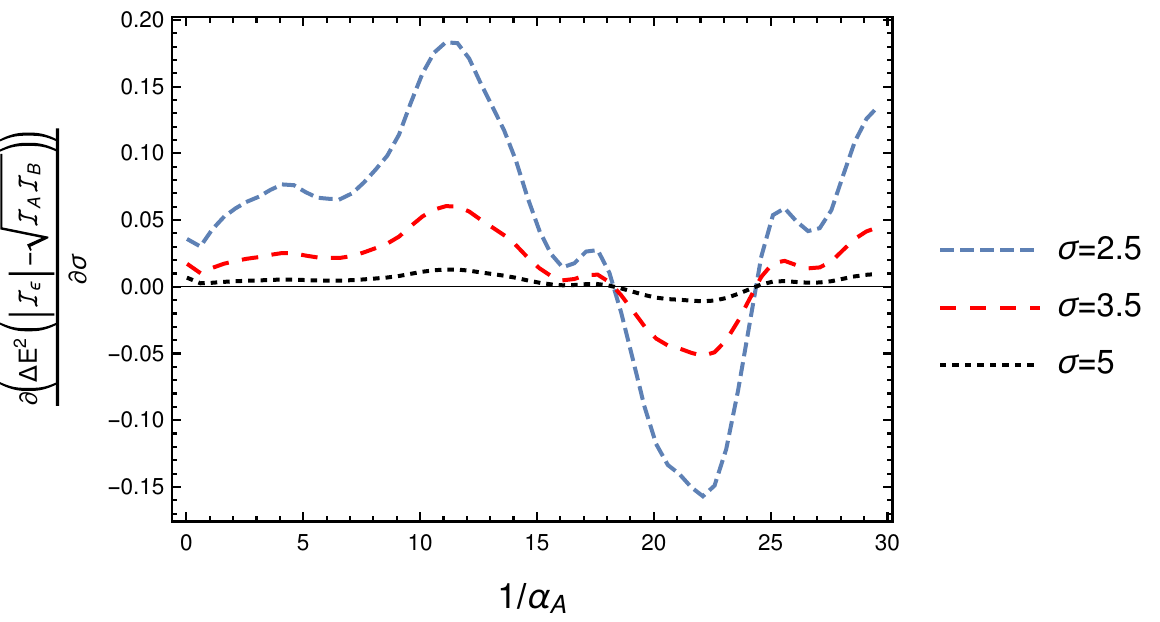}
 \caption{In $(1+3)$ dimensions the derivative with respect to $\sigma$ of the 
quantity $\Delta E^{2}\left(|\mathcal{I}_{\varepsilon}|- \sqrt{\mathcal{I}_{A} 
\mathcal{I}_{B}}\right)$ is plotted for two anti-parallelly accelerating 
detectors for varying acceleration of the first detector $\alpha_{A}$. The 
temperatures of the thermal bath $\sigma$ and other parameter $\alpha_{B}=1$ 
are 
fixed.}
\label{fig:DEHC-1p3-ab1-Va}
\end{figure}

Like the $(1+1)$ dimensional case here also we observe that when $\Delta 
E^{B}\neq \Delta E^{A}$ the integral $\mathcal{I}^{R}_{ \varepsilon } $ becomes 
finite, whereas $\mathcal{I}^{W}_{ \varepsilon } $ vanishes. On the other hand, 
from (\ref{eq:Ij-evaluated-1p3}) we observe that the integrals $\mathcal{I}_{j}$ 
have a multiplicative $\delta(0)$ term in them. Then in this situation one 
cannot harvest any entanglement. Entanglement harvesting may become possible 
only when $\Delta E^{B}=\Delta E^{A}$. In that case we consider $\Delta 
E^{B}=\Delta E^{A}=\Delta E$, which also results in $\Delta \widetilde{E}=\Delta 
E$. Then from Eq. (\ref{eq:IE-1p3-1}) and (\ref{eq:IE-1p3-2d}) one can obtain 
the expression of the integral $\mathcal{I}_{\varepsilon}$ corresponding to two 
anti-parallelly accelerated observers in $(1+3)$ dimensions as 
$\mathcal{I}_{\varepsilon}= \mathcal{I}^{W}_{\varepsilon}+\mathcal{I}^{R}_{ 
\varepsilon } $, and then get the condition for entanglement harvesting 
(\ref{eq:cond-entanglement}) to be
\begin{eqnarray}\label{eq:Cond-EntHarvest-ab-1p3}
&& \scalebox{1.1}{$\Bigg(\frac{e^{-\frac{\pi\Delta 
E}{a_{A}}}}{1-e^{-\beta\Delta E}}+\frac{e^{\frac{\pi\Delta 
E}{a_{A}}}}{e^{\beta\Delta E}-1}\Bigg)  \Bigg(\frac{e^{-\frac{\pi\Delta 
E}{a_{B}}}}{1-e^{-\beta\Delta E}}+\frac{e^{\frac{\pi\Delta 
E}{a_{B}}}}{e^{\beta\Delta E}-1}\Bigg)$} \nonumber\\
~&&~~~~~~\times~\Upsilon\left(\Delta 
E,a_{A},a_{A}\right)\Upsilon\left(\Delta 
E,a_{B},a_{B}\right)<\nonumber\\
~&& ~~~~~~~~~4\scalebox{1.1}{$\Bigg[\frac{e^{\frac{\pi\Delta 
E}{2}\left(\frac{1}{a_{B}}-\frac{1}{a_{A}}\right)}}{1-e^{-\beta\Delta 
E}}+\frac{e^{-\frac{\pi\Delta 
E}{2}\left(\frac{1}{a_{B}}-\frac{1}{a_{A}}\right)}}{e^{\beta\Delta 
E}-1}$}\nonumber\\
~&& ~~~~~~~~~~~\scalebox{0.97}{$-\sinh{\left\{\frac{\pi\Delta 
E}{2}\left(\frac{1}{a_{B}}-\frac{1}{a_{A}}\right)\right\}}\Bigg]^2 
\Upsilon\left(\Delta E,a_{A},a_{B}\right)^2$}.\nonumber\\
\end{eqnarray}

\underline{\it Numerical analysis:} --
In Fig. \ref{fig:EHC-1p3-ab1-Va} we have plotted the quantity 
$\mathcal{C}_{\mathcal{I}}=\Delta E^2\left(|\mathcal{I}_{\varepsilon}|- 
\sqrt{\mathcal{I}_{A} \mathcal{I}_{B}}\right)$, which signifies the concurrence, 
with respect to $\alpha_{A}$ for different fixed temperature of the thermal 
bath. Like the $(1+1)$ dimensional case here also we have removed the 
$\delta(0)$ factor from $\mathcal{C}_{\mathcal{I}}$, which now describes a rate 
of concurrence per unit proper time. On the other hand, in Fig. 
\ref{fig:EHC-1p3-ab1-Vbeta} and Fig. \ref{fig:EHC-1p3-ab1-Vbeta2} we have 
plotted this $\mathcal{C}_{\mathcal{I}}$ with respect to the inverse temperature 
of the thermal bath $\sigma=\beta\Delta E$ for different fixed $\alpha_{A}$. 
From both of these figures we observe that higher temperature of the thermal 
bath results in a failure of the condition for entanglement harvesting for 
accelerations much lower than the critical acceleration, which is in agreement 
with the understandings gained from the $(1+1)$ dimensional analysis. However, 
the characteristics of the curves obtained from Fig. \ref{fig:EHC-1p3-ab1-Va} 
are turbulent compared to the $(1+1)$ dimensional curves of Fig. 
\ref{fig:EHC-1p1-ab1-Va} in similar situation. It is also noticed that unlike 
the $(1+1)$ dimensional case there are multiple transition points of 
$\alpha_{A}$ in curves of Fig. \ref{fig:EHC-1p3-ab1-Va}. After crossing each of 
these transition points the characteristics of $\mathcal{C}_{\mathcal{I}}$ flips 
with respect to $\beta$, i.e., in some of the regions, in between these 
transition points, $\mathcal{C}_{\mathcal{I}}$ increases with increasing 
$\beta$, and in the neighboring regions $\mathcal{C}_{\mathcal{I}}$ decreases 
with increasing $\beta$. In Fig. \ref{fig:DEHC-1p3-ab-Vbeta} and 
\ref{fig:DEHC-1p3-ab1-Va} we have plotted the derivative of 
$\mathcal{C}_{\mathcal{I}}$ with respect to $\sigma$ for varying $\sigma$ and 
$\alpha_{A}$ to further confirm the positions of the transition points. Another 
intriguing thing to notice is that in $(1+1)$ dimensions for 
$\alpha_{A}\neq\alpha_{B}$ we observed that for a fixed temperature of the 
thermal field entanglement harvesting is possible for any accelerations above a 
certain acceleration. However, here in $(1+3)$ dimensions this is not the case, 
as now entanglement harvesting is possible in discrete ranges of $\alpha_{A}$ 
for certain values of fixed temperatures of the thermal fields.

\begin{figure}[h]
\centering
 \includegraphics[width=1.05\linewidth]{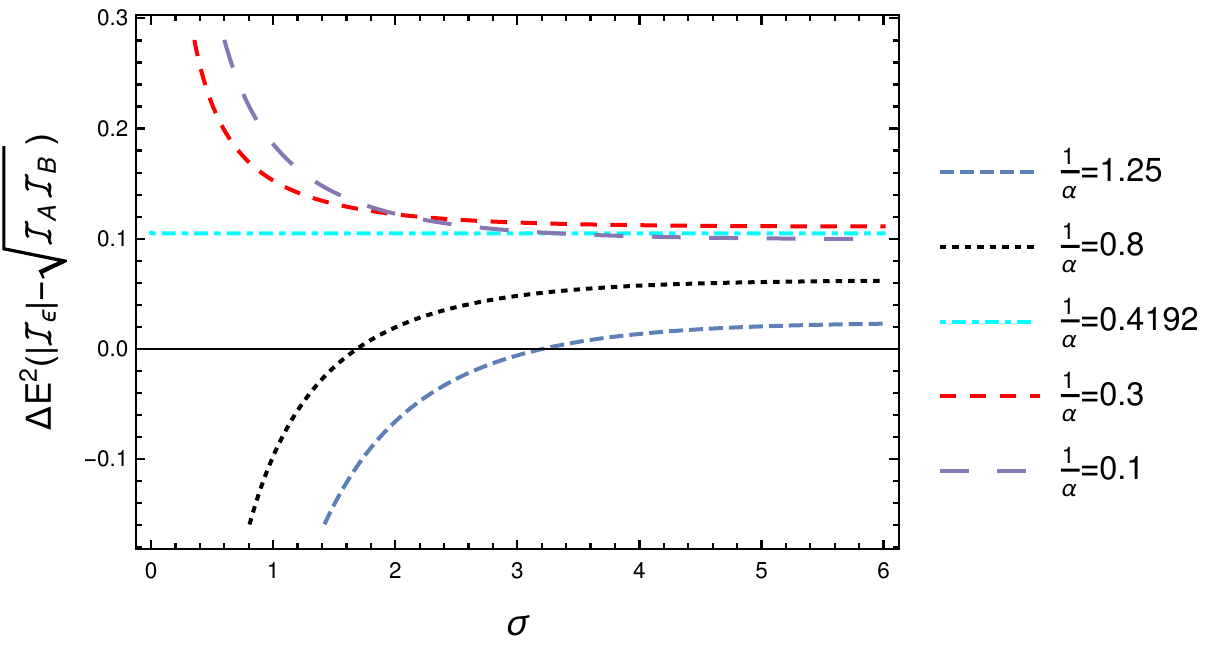}
 \caption{In $(1+3)$ dimensions the quantity $\Delta 
E^2\left(|\mathcal{I}_{\varepsilon}|- \sqrt{\mathcal{I}_{A} 
\mathcal{I}_{B}}\right)$ is plotted for two anti-parallelly accelerating 
detectors with respect to the inverse temperature of the thermal bath $\sigma$ 
for equal magnitude of proper accelerations, i.e., $\alpha_{A}=\alpha_{B}$.} 
\label{fig:EHC-1p3-aa-Vbeta}
\end{figure}

\begin{figure}[h]
\centering
 \includegraphics[width=1.05\linewidth]{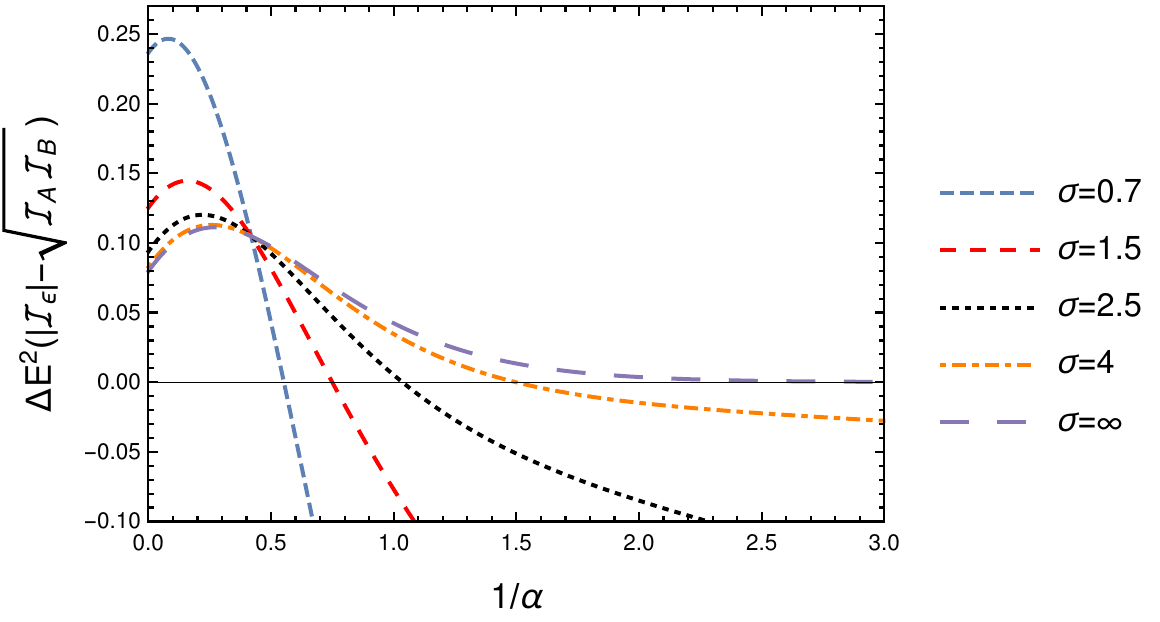}
 \caption{In $(1+3)$ dimensions the quantity $\Delta 
E^2\left(|\mathcal{I}_{\varepsilon}|- \sqrt{\mathcal{I}_{A} 
\mathcal{I}_{B}}\right)$ is plotted for two anti-parallelly accelerating 
detectors with respect to $1/\alpha$ for fixed inverse temperatures of the 
thermal bath $\sigma$ for equal magnitude of proper accelerations, i.e., 
$\alpha_{A}=\alpha_{B}$.} 
\label{fig:EHC-1p3-aa-Va}
\end{figure}

\begin{figure}[h]
\centering
 \includegraphics[width=1.05\linewidth]{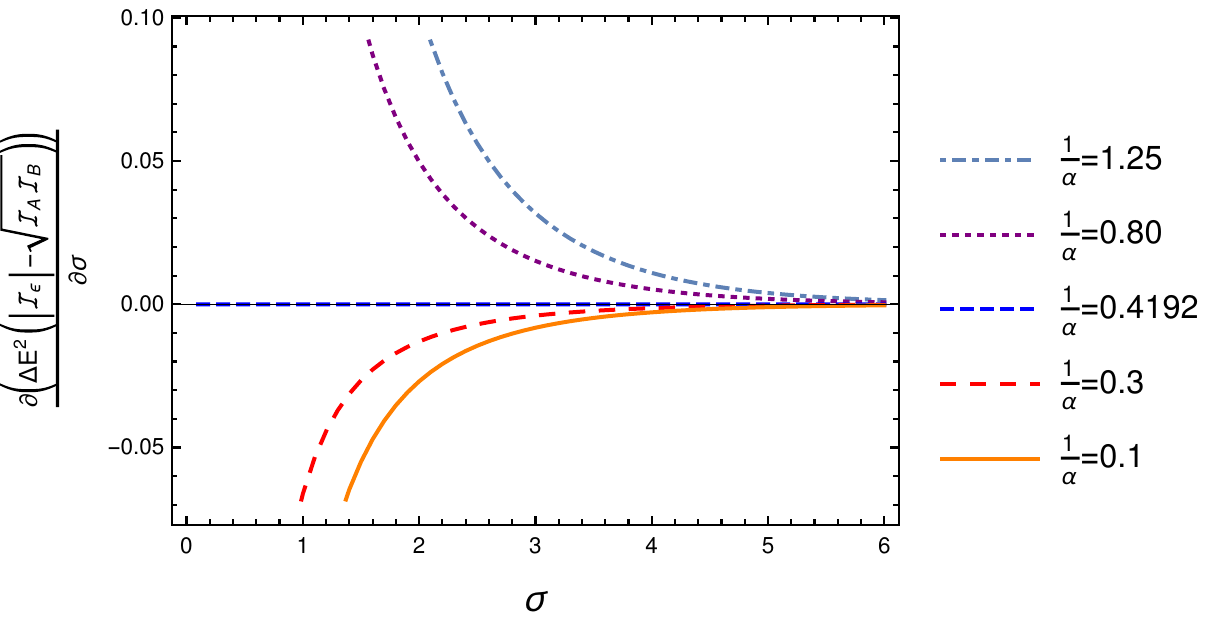}
 \caption{In $(1+3)$ dimensions the quantity $(\partial/\partial\sigma) 
\left(\Delta E^2\left(|\mathcal{I}_{\varepsilon}|- \sqrt{\mathcal{I}_{A} 
\mathcal{I}_{B}}\right)\right)$ is plotted for two anti-parallelly accelerating 
detectors with respect to the inverse temperature of the thermal bath $\sigma$ 
for equal magnitude of proper accelerations $\alpha_{A}=\alpha_{B}$.}
\label{fig:DEHC-1p3-aa-Vbeta}
\end{figure}

\begin{figure}[h]
\centering
 \includegraphics[width=1.05\linewidth]{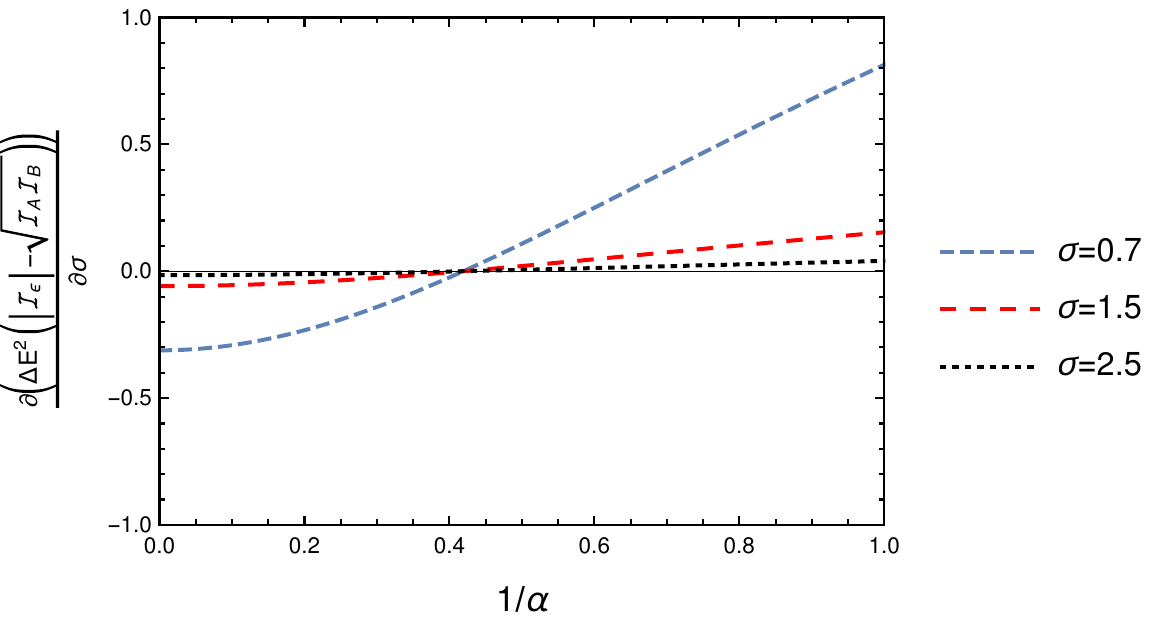}
 \caption{In $(1+3)$ dimensions the quantity $(\partial/\partial\sigma) 
\left(\Delta E^2\left(|\mathcal{I}_{\varepsilon}|- \sqrt{\mathcal{I}_{A} 
\mathcal{I}_{B}}\right)\right)$ is plotted for two anti-parallelly accelerating 
detectors with respect to the equal magnitude of proper accelerations 
$\alpha_{A}=\alpha_{B}$ for fixed inverse temperature of the thermal bath 
$\sigma$.}
\label{fig:DEHC-1p3-aa-Va}
\end{figure}

It is to be noted that in the equal magnitude of acceleration limit the second 
integral from Eq. (\ref{eq:IE-1p3-2d}) coming from the retarded Green's 
function 
vanishes and one is left with only $\mathcal{I}_{\varepsilon}= 
\mathcal{I}^{W}_{\varepsilon}$. In this particular case $a_{A}=a_{B}=a$, and 
the 
condition for entanglement harvesting from Eq. 
(\ref{eq:Cond-EntHarvest-ab-1p3}) 
becomes same as the one from the $(1+1)$ dimensional case of Eq. 
(\ref{eq:Cond-EntHarvest-aa-1p1}). Then it is expected that the entanglement 
measure $\mathcal{C}_{\mathcal{I}}$ in $(1+3)$ should be qualitatively same as 
the one from $(1+1)$ dimensions. However, it is quantitatively different in the 
$(1+3)$ dimensional case compared to the $(1+1)$ dimensional case with equal 
acceleration. In Fig. \ref{fig:EHC-1p3-aa-Vbeta} and \ref{fig:EHC-1p3-aa-Va} we 
have further plotted this quantity $\mathcal{C}_{\mathcal{I}}$ signifying the 
concurrence, in this case in $(1+3)$ dimensions. Here also the concurrence 
shows similar characteristics as was observed in the $(1+1)$ dimensional case. 
From Fig. \ref{fig:EHC-1p3-aa-Va} it is clear that the temperature of the 
thermal bath diminishes the range of acceleration in which entanglement 
extraction is possible. However, it enhances the amount of concurrence above a 
certain value of acceleration thus enhancing the entanglement extraction in 
that 
region. Furthermore, in Fig. \ref{fig:DEHC-1p3-aa-Vbeta} and 
\ref{fig:DEHC-1p3-aa-Va} we have plotted the derivative of 
$\mathcal{C}_{\mathcal{I}}$ with respect to $\sigma$ in this case for varying 
$\sigma$ and $\alpha$ for the perception of $\alpha_{c}$. It should be noted 
that in $(1+3)$ dimensions one is left out with only one transition point, 
contrary to multiple transition points in $\alpha_{A}$ from Fig. 
\ref{fig:EHC-1p3-ab1-Va}, when equal accelerations are considered.

\section{Mutual information}\label{Sec:MutualInf}

From Eq. (\ref{eq:MI-explicit}) and (\ref{eq:P-pm}) it is 
observed that the mutual information corresponding to the two accelerated 
detectors interacting with background thermal field can be estimated by 
estimating the quantities $P_{j}$ and $P_{AB}$. From Eq. 
(\ref{eq:Ij-evaluated-RRW-1p1}) and (\ref{eq:Ij-evaluated-1p3}) one can find out 
the expressions of $P_{j}$ in $(1+1)$ and $(1+3)$ dimensions corresponding to 
observers accelerated parallelly or anti-parallelly. Then here we only have to 
find out the expression of $P_{AB}$ to understand the nature of the mutual 
information for the considered detector pair. In particular we are going to 
estimate $\mathcal{I}_{AB}$ of (\ref{eq:all-integrals}) from which it is 
straightforward to estimate $P_{AB}$ using Eq. (\ref{eq:all-PJs}). We shall 
first consider the prallelly and then anti-parallelly accelerated detectors to 
estimate these quantities.

\subsection{Parallel acceleration}

\subsubsection{$(1+1)$ dimensions}

We consider the Wightman function of Eq. 
(\ref{eq:Greens-fn-TU-RRW}) corresponding to parallelly accelerated detectors 
interacting with thermal fields, and consider a change of variables $\tilde{v} = 
\tau_{B} + \tau_{A}$ and $\tilde{u}=\tau_{B}-\tau_{A}$ to evaluate the integral 
$\mathcal{I}_{AB}$ as
\begin{eqnarray}\label{eq:IAB-parallel-1p1}
 && \mathcal{I}_{AB} = \int_{-\infty}^{\infty}d\tau_{B} 
\int_{-\infty}^{\infty}d\tau_{A}~\scalebox{0.91}{$e^{i(\Delta 
E^{A}\tau_{A}-\Delta E^{B}\tau_{B})} $}\times\nonumber\\
&& ~~~~~~~~~~~~~~~~~~~~~~~~~~G^{\beta}_{W_{R}}(X_{B},X_{A})\nonumber\\
&=& \delta\left(\frac{\Delta E^A-\Delta E^B}{\sqrt{a_Aa_B}}\right) 
\frac{\pi}{\Delta \widetilde{E}\sqrt{a_Aa_B}}\frac{1}{\sqrt{\sinh{\frac{ 
\pi\Delta\widetilde{E}}{ a_B}}\sinh{\frac{\pi\Delta\widetilde{E}}{a_A}}} 
}\nonumber\\
&&~~~~~~~~~~\left[\frac{e^{-\frac{\pi\Delta\widetilde{E}}{2}\left(\frac{1 } 
{a_{B}}+\frac{1}{a_{A}}\right)}}{1-e^{-\beta\Delta\widetilde{E} 
}}+\frac{e^{\frac{\pi\Delta\widetilde{E}}{2}\left(\frac{1}{a_{B}}+\frac{1}{a_{A 
}} \right)}}{e^{\beta\Delta\widetilde{E}}-1}\right]~,
\end{eqnarray}
where $\Delta \widetilde{E} = (\Delta E^{B} + \Delta E^{A})/2$. 
It is to be noted that when $\Delta E^{B} \neq \Delta E^{A}$, due to the Dirac 
delta distribution $\delta\left((\Delta E^A-\Delta E^B)/\sqrt{a_Aa_B}\right)$ 
in front of the expression (\ref{eq:IAB-parallel-1p1}), the quantity 
$\mathcal{I}_{AB}$ (i.e., $P_{AB}$) vanishes. Then one can observe from Eq. 
(\ref{eq:P-pm}) that the quantities $P_{\pm}$ become $P_{A}$ and $P_{B}$, which 
in turn leads to zero value of the mutual information from 
(\ref{eq:MI-explicit}).

\begin{figure}[h]
\centering
 \includegraphics[width=1.0\linewidth]{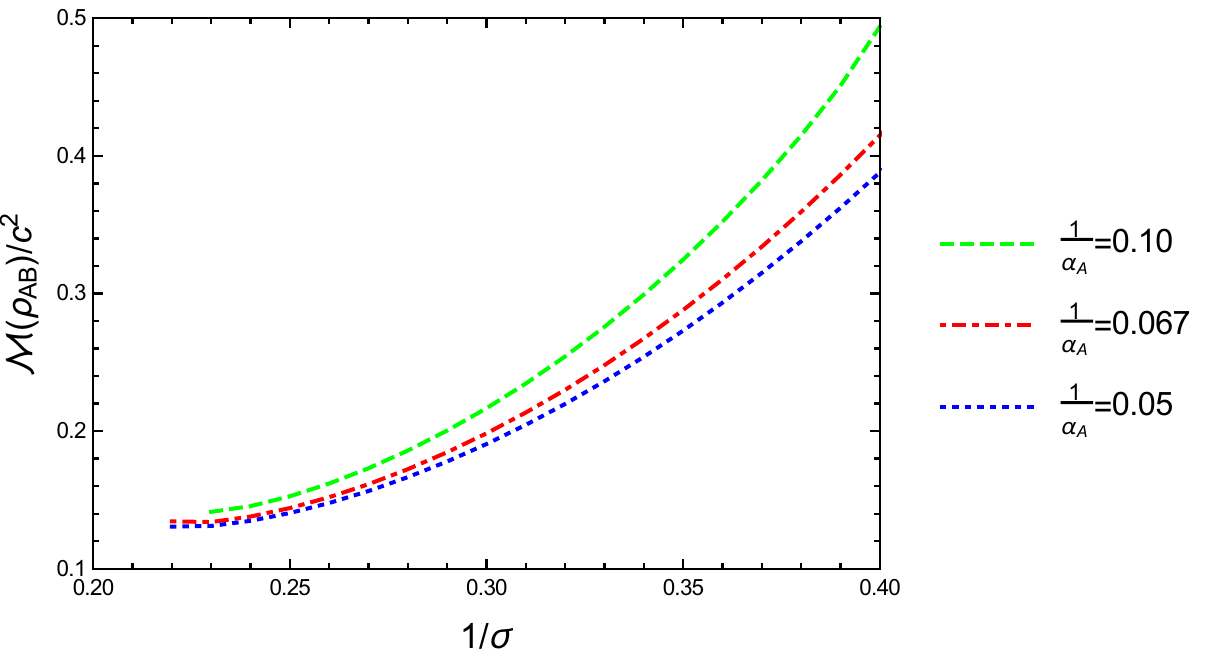}
 \caption{In $(1+1)$ dimensions the quantity $\mathcal{M}(\rho_{AB})/c^2$ per 
unit proper time is plotted, which signifies the mutual information, for two 
parallelly accelerating detectors with respect to the temperature of the thermal 
field $T^{(f)}(\sim 1/\sigma)$ for different fixed proper accelerations 
$\alpha_{A}$, where $\alpha_{B}=1$.}
\label{fig:MI-1p1-ab-a-Vbeta}
\end{figure}

One has non vanishing mutual information only when $P_{AB}\neq 
0$, i.e., when $\Delta E^{B} = \Delta E^{A}$. We get $\Delta \widetilde{E} = 
\Delta E$ by considering $\Delta E^{B} = \Delta E^{A} =\Delta E$. In that case 
it is observed that there will be a multiplicative $\delta(0)$ term in the 
expression of $\mathcal{I}_{AB}$ similar to the case of $\mathcal{I}_{j}$ of 
(\ref{eq:Ij-evaluated-RRW-1p1}). One can remove this $\delta(0)$ term attributed 
to considering a rate per unit proper time of $\mathcal{I}_{AB}$. On the other 
hand, from (\ref{eq:all-PJs}) it is observed that for the exact evaluation of 
$P_{A}$, $P_{B}$, and $P_{AB}$ it is imperative to know the expectation value 
$\langle E_{1}^{j}|m_{j}(0)|E_{0}^{j}\rangle$, which can be estimated for an explicit choice of the 
monopole operator $m_{j}(0) = |E_{1}^{j}\rangle 
\langle E_{0}^{j}| + |E_{0}^{j}\rangle \langle E_{1}^{j}|$.
It is to be noted that in the expression of the concurrence from 
(\ref{eq:concurrence-gen-exp}) there was a common multiplicative term $|\langle 
E_{1}^{B}|m_{B}(0)| E_{0}^{B}\rangle| |\langle E_{1}^{A}|m_{A}(0)| 
E_{0}^{A}\rangle|$, which we neglected concentrating only on the effect of the 
spacetime on detector response. However, for the case of the mutual information 
of (\ref{eq:MI-explicit}) one cannot pull out a common multiplicative 
expectation of the monopole operator and we have to explicitly put their values 
for a numerical evaluation.
In particular, for both $j=A$ and $j=B$ it is observed that $\langle E_{1}^{j} 
|m_{j}(0)| E_{0}^{j}\rangle=1$. Then using Eq. (\ref{eq:all-PJs}), 
(\ref{eq:all-integrals}), (\ref{eq:MI-explicit}), and (\ref{eq:P-pm}) one can 
explicitly evaluate the mutual information in this case. In Fig. 
\ref{fig:MI-1p1-ab-a-Vbeta} we have plotted the rate of mutual information with 
respect to the temperature of the thermal field $T^{(f)}$ ($\sim 1/\sigma = 
1/(\beta\Delta E)$), which shows that with increasing temperature the mutual 
information increases. From this figure it is also observed that with 
increasing acceleration of the first detector (signified by $\alpha_{A}$) the 
mutual information decreases.

\subsubsection{$(1+3)$ dimensions}

We consider the positive frequency Wightman function (\ref{eq:Greens-fn-TU-3D}) 
for the estimation of the quantity $\mathcal{I}_{AB}$ in $(1+3)$ dimensions, 
which becomes
\begin{eqnarray}\label{eq:IAB-parallel-1p3}
 && \mathcal{I}_{AB} = \int_{-\infty}^{\infty}d\tau_{B} 
\int_{-\infty}^{\infty}d\tau_{A}~\scalebox{0.91}{$e^{i(\Delta 
E^{A}\tau_{A}-\Delta E^{B}\tau_{B})} $}\nonumber\\
&& ~~~~~~~~~~~~~~~~~~~~~~~~~~G^{\beta}_{W^{3D}_{R}}(X_{B},X_{A})\nonumber\\
&=& \delta\left(\frac{\Delta E^A-\Delta 
E^B}{\sqrt{a_Aa_B}}\right) 
\frac{1}{\pi a_A a_B} 
\Upsilon\left(\Delta\widetilde{E},a_{A},a_{B}\right)\nonumber\\
&&~~~~~~\left[\frac{e^{-\frac{\pi\Delta\widetilde{E}}{2}\left(\frac{1} 
{a_{B}}+\frac{1}{a_{A}}\right)}}{1-e^{-\beta\Delta\widetilde{E} 
}}+\frac{e^{\frac{\pi\Delta\widetilde{E}}{2}\left(\frac{1}{a_{B}}+\frac{1}{a_{A 
}} \right)}}{e^{\beta\Delta\widetilde{E}}-1}\right].
\end{eqnarray}
Here also $\Delta \widetilde{E} = (\Delta E^{B} + \Delta 
E^{A})/2$, and for $\Delta E^{B} \neq \Delta E^{A}$ Dirac delta distribution 
$\delta\left((\Delta E^A-\Delta E^B)/\sqrt{a_Aa_B}\right)$ in 
(\ref{eq:IAB-parallel-1p3}) provides vanishing $\mathcal{I}_{AB}$ (or $P_{AB}$). 
This leads to vanishing mutual information.

\begin{figure}[h]
\centering
 \includegraphics[width=1.0\linewidth]{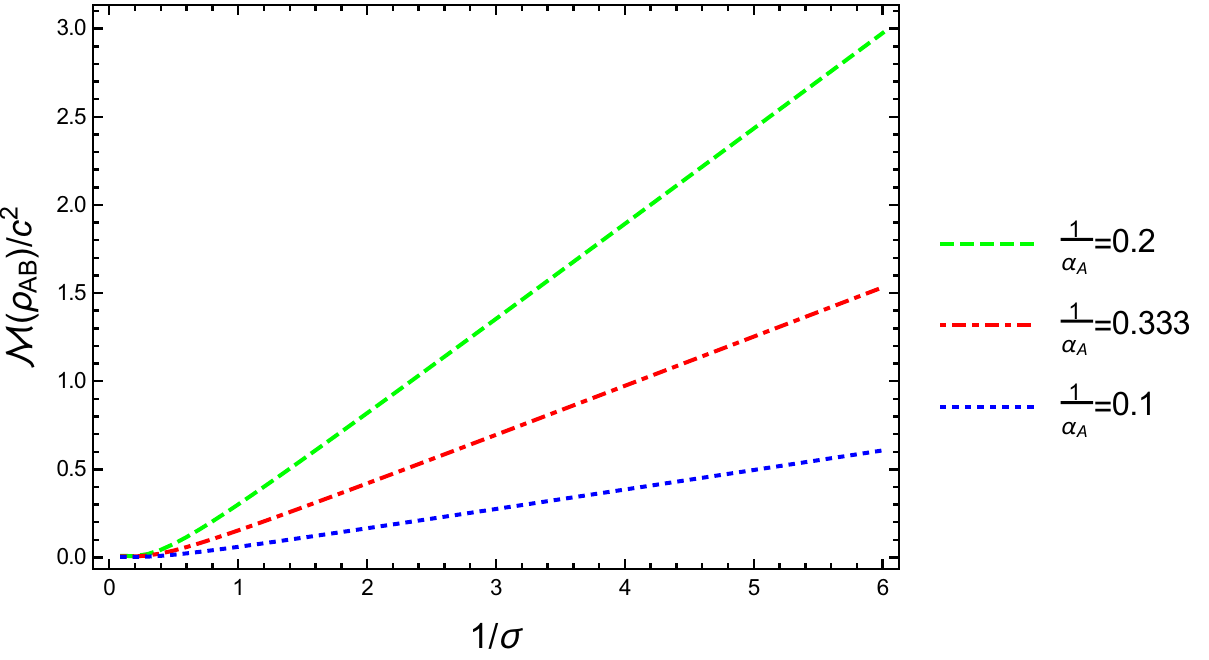}
 \caption{In $(1+3)$ dimensions the quantity $\mathcal{M}(\rho_{AB})/c^2$ per 
unit proper time is plotted, which signifies the mutual information, for two 
parallelly accelerating detectors with respect to the temperature of the thermal 
field $T^{(f)}(\sim 1/\sigma)$ for different fixed proper accelerations 
$\alpha_{A}$, where $\alpha_{B}=1$.}
\label{fig:MI-1p3-ab-a-Vbeta}
\end{figure}

The mutual information is non vanishing only when $\Delta E^{B} 
= \Delta E^{A}$. Here also considering $\langle E_{1}^{j} |m_{j}(0)| 
E_{0}^{j}\rangle=1$ we have estimated the mutual information using the Eq. 
(\ref{eq:all-PJs}), (\ref{eq:all-integrals}), (\ref{eq:MI-explicit}), and 
(\ref{eq:P-pm}), and we plotted the rate of mutual information with respect to 
the temperature of the thermal field $T^{(f)}$ ($\sim1/\sigma$) in Fig. 
\ref{fig:MI-1p3-ab-a-Vbeta}. From this figure we conclude that in $(1+3)$ 
dimensions also the mutual information increases with increasing temperature of 
the thermal field and decreasing acceleration of the first detector (signified 
by $\alpha_{A}$).

\subsection{Anti-parallel acceleration}

\subsubsection{$(1+1)$ dimensions}

We consider the Wightman function from Eq. (\ref{eq:Greens-fn-TU-LR}) 
corresponding to two anti-parallelly accelerated observers in $(1+1)$ 
dimensional thermal bath, and a change of variables 
$\tilde{v}=\tau_{B}+\tau_{A}$ and $\tilde{u}=\tau_{B}-\tau_{A}$ to evaluate the 
quantity $\mathcal{I}_{AB}$ from (\ref{eq:all-integrals}). One can express this 
integral $\mathcal{I}_{AB}$ as
\begin{eqnarray}\label{eq:IAB-anti-parallel-1p1}
 && \mathcal{I}_{AB} = \int_{-\infty}^{\infty}d\tau_{B} 
\int_{-\infty}^{\infty}d\tau_{A}~\scalebox{0.91}{$e^{i(\Delta 
E^{A}\tau_{A}-\Delta E^{B}\tau_{B})} $}\times\nonumber\\
&& ~~~~~~~~~~~~~~~~~~~~~~~~~~G^{\beta}_{W_{LR}}(X_{B},X_{A})\nonumber\\
&=& \delta\left(\frac{\Delta E^A+\Delta E^B}{\sqrt{a_Aa_B}}\right) 
\frac{\pi}{\Delta \breve{E}\sqrt{a_Aa_B}}\frac{1}{\sqrt{\sinh{\frac{ 
\pi\Delta\breve{E}}{ a_B}}\sinh{\frac{\pi\Delta\breve{E}}{a_A}}} 
}\nonumber\\
&&~~~~~~~~~~\left[\frac{e^{-\frac{\pi\Delta\breve{E}}{2}\left(\frac{1 } 
{a_{B}}-\frac{1}{a_{A}}\right)}}{1-e^{-\beta\Delta\breve{E} 
}}+\frac{e^{\frac{\pi\Delta\breve{E}}{2}\left(\frac{1}{a_{B}}-\frac{1}{a_{A 
}} \right)}}{e^{\beta\Delta\breve{E}}-1}\right]~,
\end{eqnarray}
where $\Delta \breve{E} = (\Delta E^{B} - \Delta E^{A})/2$. 
Then it is obvious that for $\Delta E^{A}>0$ and $\Delta E^{B}>0$ the Dirac 
delta distribution sitting in front of this expression $\delta\left((\Delta 
E^A+\Delta E^B)/\sqrt{a_Aa_B}\right)$ will provide a vanishing contribution. 
Thus $\mathcal{I}_{AB}$ vanishes and so vanishes $P_{AB}$. Then from Eq. 
(\ref{eq:P-pm}) one can estimate the quantities $P_{\pm}$ to be $P_{A}$ and 
$P_{B}$, which in turn leads to the expression of mutual information from 
(\ref{eq:MI-explicit}) to be vanishing upto $\mathcal{O}(c^2)$. This result 
persuades one to conclude that the mutual information corresponding to two 
anti-parallelly accelerated detectors in a thermal bath is zero in $(1+1)$ 
dimensions.

\subsubsection{$(1+3)$ dimensions}

We consider the positive frequency Wightman function 
(\ref{eq:Greens-fn-TU-3D-LR}) corresponding to anti-parallelly accelerated 
observers for the estimation of the quantity $\mathcal{I}_{AB}$ in $(1+3)$ 
dimensions, which becomes
\begin{eqnarray}\label{eq:IAB-anti-parallel-1p3}
 && \mathcal{I}_{AB} = \int_{-\infty}^{\infty}d\tau_{B} 
\int_{-\infty}^{\infty}d\tau_{A}~\scalebox{0.91}{$e^{i(\Delta 
E^{A}\tau_{A}-\Delta E^{B}\tau_{B})} $}\nonumber\\
&& ~~~~~~~~~~~~~~~~~~~~~~~~~~G^{\beta}_{W^{3D}_{LR}}(X_{B},X_{A})\nonumber\\
&=& \delta\left(\frac{\Delta E^A+\Delta 
E^B}{\sqrt{a_Aa_B}}\right) 
\frac{1}{\pi a_A a_B} 
\Upsilon\left(\Delta\breve{E},a_{A},a_{B}\right)\nonumber\\
&&~~~~~~\left[\frac{e^{-\frac{\pi\Delta\breve{E}}{2}\left(\frac{1} 
{a_{B}}-\frac{1}{a_{A}}\right)}}{1-e^{-\beta\Delta\breve{E} 
}}+\frac{e^{\frac{\pi\Delta\breve{E}}{2}\left(\frac{1}{a_{B}}-\frac{1}{a_{
A 
}} \right)}}{e^{\beta\Delta\breve{E}}-1}\right].
\end{eqnarray}
Here also $\Delta \breve{E} = (\Delta E^{B} - \Delta E^{A})/2$ 
and similar to the $(1+1)$ dimensional case the Dirac delta distribution 
$\delta\left((\Delta E^A+\Delta E^B)/\sqrt{a_Aa_B}\right)$ sitting in front of 
this expression will provide a vanishing contribution. This leads to a vanishing 
$P_{AB}$ and in turn vanishing mutual information upto $\mathcal{O}(c^2)$ from 
(\ref{eq:MI-explicit}). Then in $(1+3)$ dimensions also one can conclude that 
the mutual information corresponding to two anti-parallelly accelerated 
detectors in a thermal bath is zero.

\section{Discussion}\label{discussion}

The possibility of constructing a plausible experimental setup in contact with a 
thermal bath is much higher, as, in nature, the background is thermal than a 
purely non-thermal field vacuum. Therefore it is much more relevant to 
understand realistic situations in our surroundings to study physical phenomena 
in the presence of a thermal bath or by considering the thermal fields. We have 
considered studying entanglement harvesting with two accelerated Unruh-DeWitt 
detectors interacting with a background thermal massless scalar field in this 
work. We have constructed the relevant Green's functions corresponding to 
accelerated observers in thermal bath considering the Rindler modes with the 
vacuum for the Unruh modes to avoid dealing with Wightman functions which are 
not time translational invariant. We used the prescription of 
\cite{Barman:2021oum} for constructing the Green's functions and followed the 
entanglement harvesting mechanism of articles \cite{Koga:2018the, Koga:2019fqh}. 
It is observed that for zero temperature of the thermal bath, i.e., in the limit 
$\beta\to\infty$, in both $(1+1)$ and $(1+3)$ dimensions considering the equal 
magnitude of accelerations for the two observers but moving anti-parallelly, we 
always get the condition for entanglement harvesting to be satisfied. It ensures 
that entanglement extraction is possible for any finite non-zero acceleration in 
zero temperature background, which is in fact known from the earlier works of 
\cite{Reznik:2002fz}.

Furthermore, for non-zero temperature of the thermal bath with equal magnitude 
of acceleration of the anti-parallelly accelerated observers in both $(1+1)$ and 
$(1+3)$ dimensions we get identical conditions for entanglement harvesting (Eq. 
(\ref{eq:Cond-EntHarvest-aa-1p1})). The quantity $\mathcal{C}_{\mathcal{I}}$ 
signifying concurrence also shows similar behavior in $(1+1)$ and $(1+3)$ 
dimensions, which can be observed from Fig. \ref{fig:EHC-1p1-aa-Vbeta}, 
\ref{fig:EHC-1p1-aa-Va} and \ref{fig:EHC-1p3-aa-Vbeta}, \ref{fig:EHC-1p3-aa-Va}. 
An interesting fact we noticed from Fig. \ref{fig:EHC-1p1-aa-Va} and 
\ref{fig:EHC-1p3-aa-Va} is that with increasing temperature of the thermal bath 
(decreasing $\beta$ or $\sigma$), the range of acceleration, in which 
entanglement can be harvested, is decreasing, which is in agreement with the 
results of previous works \cite{Brown:2013kia, Simidzija:2018ddw}. We observe 
that higher acceleration is needed to initiate entanglement harvesting with the 
higher temperature of the thermal bath. However, once for a certain temperature, 
entanglement harvesting starts with some initial acceleration in this system; it 
keeps on harvesting for all other higher accelerations. On the other hand, above 
a certain critical acceleration $a=a_{c}$ we see the amount of entanglement 
harvested, denoted by concurrence, to be increasing with increasing temperature 
of the thermal bath, showing a characteristic opposite compared to the region 
below $a=a_{c}$, which is like a phase transition.

We also observe in $(1+1)$ dimensions from Eq. (\ref{eq:IE-1p1-1}) and 
(\ref{eq:IE-1p1-2d}) that for $a_{A}= a_{B}=a$ and in the limit of $a\to0$ the 
whole quantity $\mathcal{I}_{\varepsilon}$ without the multiplicative delta 
distribution vanishes, making the condition for entanglement extraction to break 
down. Then it is obvious that in the requirement of entanglement harvesting, an 
accelerated observer and a static observer in a thermal bath do not act in equal 
footing.

On the other hand, from earlier research works, it was known that the integral, 
representing the correlation between the two-detectors and responsible for 
mathematically realizing the entanglement harvesting, is related to the Wightman 
function $G_{W}(X'_{B}, X_{A})$ between two detector events.  However, the 
recent investigations \cite{Koga:2018the, Ng:2018ilp, Koga:2019fqh} suggest that 
this integral is related to the Feynman propagator $G_{F}(X'_{B}, X_{A})$. In 
particular, we observed that this additional contribution could be identified to 
be dependent on the Retarded Green's function. In our case we observed a finite 
contribution from the Retarded Green's function $G_{R}(X'_{B},X_{A})$. In 
condition for entanglement extraction from Eq. (\ref{eq:Cond-EntHarvest-ab-1p1}) 
and (\ref{eq:Cond-EntHarvest-ab-1p3}) we observed that the contribution from the 
Retarded Green's function can be identified through a quantity of 
$\sinh{\left\{\pi\Delta E \left(1/a_{B} - 1/a_{A} \right)/2\right\}}$. Then for 
all $a_{A}\neq a_{B}$ this quantity has a non-zero contribution. However, when 
$a_{A}= a_{B}$ the contributions from $G_{F}(X'_{B},X_{A})$ and 
$G_{W}(X'_{B},X_{A})$ are the same. Finally it is to be noticed that when 
$a_{A}\neq a_{B}$, by observing the plots of $\mathcal{C}_{\mathcal{I}}$ with 
respect to the acceleration of the first detector, it is possible to distinguish 
between the cases of $(1+1)$ and $(1+3)$ dimensions. Notably, in $(1+1)$ 
dimensions the curves of fixed $\sigma$ (Fig. \ref{fig:EHC-1p1-ab1-Va}) shows 
monotonic nature, while in $(1+3)$ dimensions (Fig. \ref{fig:EHC-1p3-ab1-Va}) 
this is not the case with various peaks and valleys. In $(1+3)$ dimensions this 
results in multiple transition points of accelerations $a_{A}$ between which the 
nature of concurrence with respect to the temperature of the thermal bath flips 
compared to the adjacent regions, also restricting the entanglement harvesting 
to discrete ranges of acceleration $a_{A}$ for certain fixed temperatures of the 
thermal bath.

An investigation of mutual information among the detectors has also been done 
here. We found that this vanishes for the anti-parallel situation, whereas it is 
non-vanishing for the parallel case. In the latter situation, mutual information 
increases with the increase in background field temperature while decreasing 
with the first detector's proper acceleration.

We want to mention here the nature of the curves in Fig. \ref{fig:EHC-1p1-aa-Va} 
and Fig. \ref{fig:EHC-1p3-aa-Va} which we did not discuss in the main text. In 
the equal acceleration case $a_{A}= a_{B}$ in both $(1+1)$ and $(1+3)$ 
dimensions, it is observed that after a specific critical acceleration, the 
entanglement extraction rate tends to decrease with increasing acceleration. The 
possible reason can be as follows. When the acceleration of the detector is 
substantial (i.e. $a_A\rightarrow \infty$), the detector moves very near to the 
null surface denoted by $X=-T$ and $X=T$ and also feels a very high temperature 
due to its acceleration (temperature is given by Unruh expression $a/(2\pi)$). 
In this regime, the thermal bath due to acceleration becomes equally relevant 
along with the real thermal bath on the nature of entanglement harvesting.  
Since we already observed that temperature could reduce the entanglement between 
the detectors, both temperatures due to the Unruh effect and the thermal bath 
may play a role in the decreasing nature of concurrence. It is happening in a 
very high acceleration regime as there the Unruh temperature also becomes 
appreciable to affect entanglement harvesting. So in this regime, acceleration 
is showing its double standards -- on one side, it is helping in entanglement, 
but on another side, it is also suppressing this phenomenon. In lower 
accelerations, the Unruh temperature is not so palpable to affect entanglement 
harvesting. Therefore there the acceleration plays only the role in helping 
entanglement. In this regard, we point out that this reason is only a suggestive 
one, and further investigation is needed to find any conclusive explanation.

Finally, we mention that in this paper, we deeply investigated the effect of 
background temperature on entanglement harvesting between two uniformly 
accelerated detectors. As we mentioned, this situation mimics a much more 
realistic situation, and hence the results have practical importance. As we 
mentioned above, the background temperature introduces several interesting 
noticeable features absent when the temperature is zero. Therefore we feel that 
the present study is significant in entanglement harvesting between the 
observers through their interaction with the background quantum fields and helps 
in the progress of the above subject.

\begin{acknowledgments}
DB and SB would like to thank the Indian Institute of Technology Guwahati (IIT 
Guwahati) for supporting this work through Doctoral and Post-Doctoral 
Fellowships. The research of BRM is partially supported by a START-UP RESEARCH 
GRANT (No. SG/PHY/P/BRM/01) from the Indian Institute of Technology Guwahati, 
India.
\end{acknowledgments}

 

\bibliographystyle{apsrev}

\bibliography{bibtexfile}

\end{document}